\documentclass[preprint]{aastex}
\usepackage{epsf}
\usepackage{natbib}
\usepackage{subfigure}

\newcommand{\be}{\begin{equation}}
\newcommand{\ee}{\end{equation}}

\newcommand{\feh}{\hbox{$ [{\rm Fe}/{\rm H}]$ }} 
\def\c2{\chi ^2}

\begin{document}

\title{Metallicity Analysis of MACHO Galactic Bulge RR0 Lyrae stars from 
their Lightcurves}

\author{Andrea Kunder and Brian Chaboyer} 
\affil{Dartmouth College, 6127 Wilder Lab, Hanover, NH 03755}
\affil{E-mail: Andrea.M.Kunder@Dartmouth.edu and Brian.Chaboyer@dartmouth.edu}


\begin{abstract}
We present metallicities of 2690 RR0 Lyrae stars observed toward the 
MACHO Survey fields in the Galactic Bulge. These \feh values
are based upon an empirically calibrated relationship that uses
the Fourier coefficients of the light curve and are accurate to
$\pm$ 0.2 dex.  The majority of the RR0 Lyrae stars in our sample are 
located in the Galactic Bulge, but 255 RR0 stars 
are associated with Sagittarius dwarf galaxy.
We find that the RR0 Lyrae stars that belong to the Galactic Bulge
have average metallicities $\feh = -$1.25, with a broad 
metallicity range from $\feh = -$2.26 to $-0.15$.
The RR0 stars from Sagittarius dwarf galaxy have lower average metallicity 
of $\feh = -1.55 \pm 0.02$, with an intrinsic 
dispersion of 0.25 dex, similar to that in the bulge.  A 
correlation between metallicity and galactocentric distance is found, 
in a sense that for the metal-poor RR0 Lyrae stars (\feh $< -$1.5 dex), 
the closer a star is to the Galactic center, on average, 
the more metal rich it is.  However, for the metal-rich RR0 Lyrae stars 
(\feh $> -$1.2 dex), this trend is reversed.
Using mean magnitudes of MACHO RR Lyrae stars, we 
searched for the evidence of the Galactic bar, and found marginal evidence
of a bar.
The absence of a strong bar indicates that the RR Lyrae in the bulge 
represent a different population than the majority of the bulge stars,
which are metal rich and are part of a bar.  
\end{abstract}

\section{Introduction}
The bulge RR Lyrae variables are likely to be among the oldest and most 
metal poor stars in the bulge, so their metal abundances are of considerable 
importance in determining the mix of populations in the Galactic Bulge, 
and vital to our understanding of the nature of the bulge itself 
\citep{walkThisTurn}.  Studying abundances of RR Lyrae stars lets us
probe early chemical evolution of the Milky Way and allows the chemical 
history of the the oldest population of the bulge to be traced out.

Direct studies of the bulge are difficult due to the severe crowding 
toward the central regions of the Galaxy and the large, patchy, reddening 
along the line of sight. Therefore, astronomical literature contains limited 
spectroscopy of RR Lyrae stars toward the bulge and most spectroscopic
studies focus on Baade's window (BW) centered roughly on 
the globular cluster NGC 6522 at $(l, b) = (1 \hbox{$.\!\!^\circ$} 0,
-3 \hbox{$.\!\!^\circ$} 9)$.  \citet{but76}
measured $\Delta S$ abundances \citep{preston59} for nine RR Lyrae stars 
in BW and derived a mean metallicity of $<\feh>$ $= -0.82 \pm 0.14 $ 
from a broad abundance distribution.  \citet{gratzi} measured $\Delta S$ 
values for 17 BW variables and found $<\feh>$ $= -0.76 \pm 0.12$.  
Again the results suggested a wide range of abundance.  
\citet{walkThisTurn} found a somewhat lower abundance of $<\feh>$ $= -1.0$. 
from 59 RR Lyrae variables in BW.
In contrast to other studies, they concluded  that the RR Lyrae stars in 
BW had a relatively narrow range in abundance.  

In this paper, we determine iron metallicities of bulge 
RR0\footnote{RR0 stars have 
been traditionally called RRab stars and are simply fundamental pulsators.  
See \citet{alc00} for further explanation on the more intuitive 
nomenclature.}  Lyrae stars determined 
indirectly from pulsational properties of the variables.  A number of 
studies have shown that Fourier parameters of light curves of RR Lyrae
stars are related to their physics properties, including 
the metallicity \citep{sc93}.  \citet{jk96} used spectroscopic and 
photometric observations of field RR0 stars to calibrate a relationship 
between the \feh of RR0 Lyrae stars and Fourier parameters.  
Because of the accurate fit of the Fourier formula to the 
observed metallicities of the RR0 Lyrae field stars, it
appears to be very attractive to use this on various large databases.
Recently, \citet{kinemuchi06} applied the \citet{jk96} formula to find
photometric metallicities of 433 of 1188 RR0 Lyrae stars in the Northern Sky
Variability Survey.  Although this \feh method has been used frequently 
and appears to be reasonably reliable, there has been some question as 
to the validity of
the \citet{jk96} method \citep[see discussions in][]{diFab05, grat04}.
This is investigated in detail in Appendix A where it is demonstrated
that the \feh derived from MACHO LMC RR0 Lyrae lightcurves using 
\citet{jk96} method agrees well with spectroscopic determinations from
\citet{grat04} and \citet{bori06}.  We utilize the Fourier coefficients for 
2690 RR0 Lyrae stars in the MACHO bulge survey to derive 
\feh abundances of these variables.  

This paper first addresses the observational data and
Fourier Coefficients in \S2.  Next, the derivation of \feh, along with
various tests, is described in \S3. 
The presentation of bulge and Sgr metallicities
and implications regarding these metallicities follow in \S4.  Using 
reddenings determined from the minimum V-band light of the RR0
Lyrae stars, in \S5, trends coinciding with distance are elucidated. 
A summary is presented in \S6.

\section{Data}

The observations of the Galactic Bulge were made by the MACHO team
with a 1.27-meter telescope at Mount Stromlo Observatory, 
Australia \citep[e.g.][]{alc97}.  In total, 7 seasons (1993-1999) 
of useful data were collected in the 94 
Galactic Bulge fields. The two pairs of four 2k x 2k CCDs produced
simultaneous imaging in non-standard blue and red passbands.  
Cook, Kunder, \& Popowski \citetext{in preparation} present 
3674 RR0 Lyrae stars toward the Galactic Bulge based on these
observations.  \citet{kunder08} use this database to find
reddenings along the line of site to 3525 RR0 Lyrae Stars.  Each
RR0 Lyrae star in their catalog has a two-filter light curve with 
several hundred photometric measurements.    

To convert the MACHO instrumental magnitudes 
$b_M$ and $r_M$ to standard Johnson's $V$ and Kron-Cousins $R$, we adopted
the following relations:
\begin{equation}
\label{vMag}
V = 23.70 + 0.82b_M + 0.18r_M,
\end{equation}
and
\begin{equation}
\label{rMag}
R = 23.41 + 0.18b_M + 0.82r_M.
\end{equation}
These are taken from \citet{alc99} under the assumptions of average airmass
and no corrections for small scale focal plane effects. Almost all 
conclusions in this paper 
are derived using Johnson $V$; only in a few cases where colors
of the RR0 Lyrae stars are needed, we refer to the Kron-Cousin $R$.

\section{Determination of the \feh metallicity}

\subsection{The method}

The phased light curve (here presented
in terms of magnitudes $m$) is represented with its $N$-th order Fourier 
decomposition:
\begin{equation}
m(f) ~=~ a_0 ~+~ \sum_{k=1}^{N} (a_k \sin(2\pi kf) ~+~ b_k \cos(2\pi kf))
\end{equation}
where $P$ is the period and $f$ is the phase
\citep[see][provides more details]{pe86, sl81, kungefe}.

\begin{figure}[htb]
\includegraphics[width=16cm]{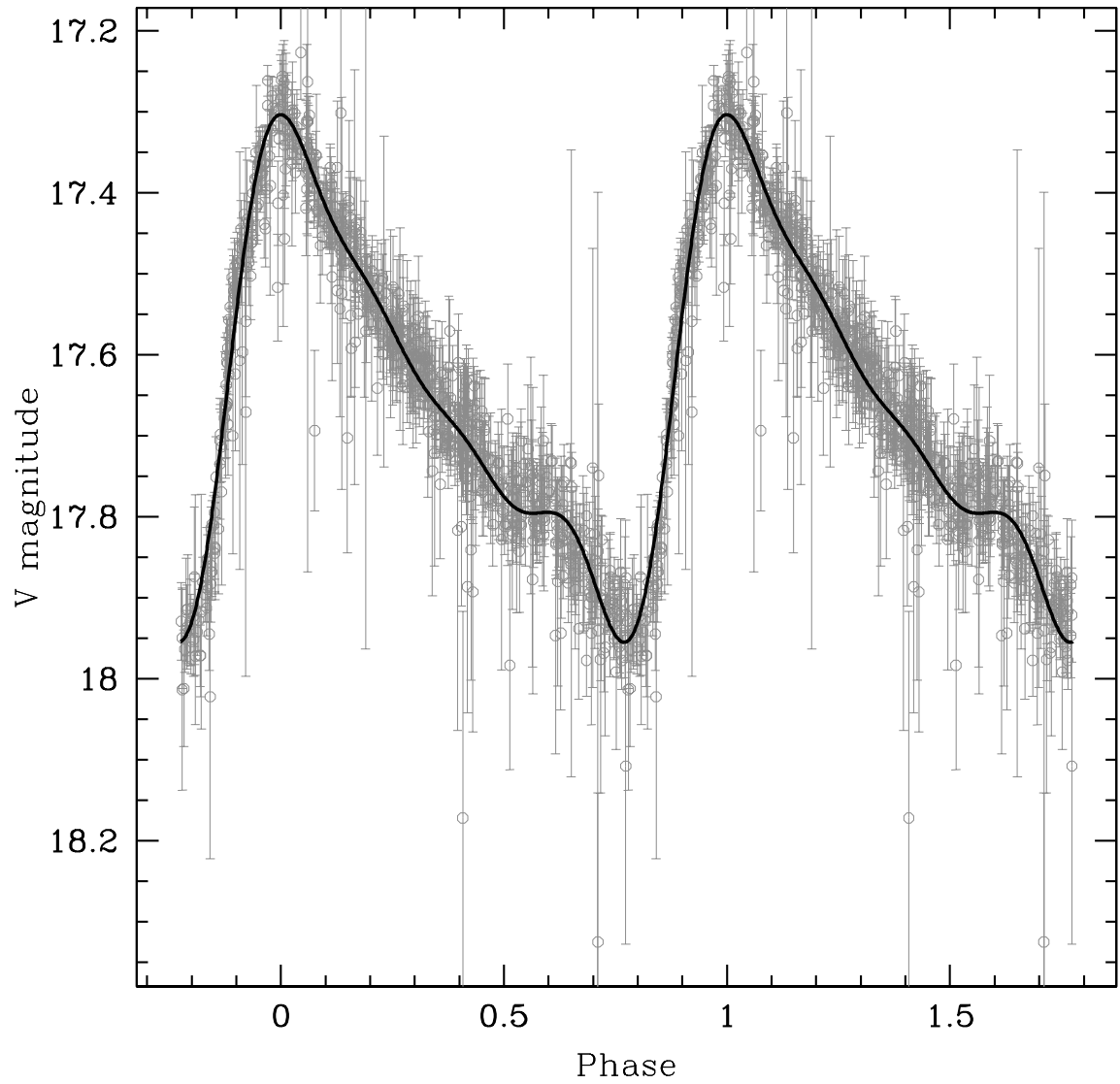}
\caption{Fourth order Fourier decomposition curve together with original
calibrated $V$-band data of a representative RR0 star 
designated 101.20778.1172 from the MACHO database.  
\label{plotone}}
\end{figure}

These coefficients are converted to phase
and amplitude parameters that coincide with a pure sine Fourier series fit,
and the \citet{jk96} period-phase-\feh relation can be applied.  
Figure~\ref{plotone} shows a representative star from our sample.  

A deviation parameter, $D_m$, defined by \citet{jk96}, represents a 
quality test on the 
regularity of the shape of the light curve.  \citet{kovkan98} have updated
values for the deviation parameters, which are employed in this
analysis.  In Appendix A, we demonstrate that a $D_m < 4.5$ and a fourth order
Fourier decomposition is best suited for the analysis of the data 
in this paper. The \citet{jk96} period-phase-\feh relation is:
\begin{equation}
\label{bigdog}
{\rm \feh_{JK}}=-5.038-5.394P + 1.345\phi_{31}
\end{equation}
where $P$ is the period, and $\phi_{31}$ is the Fourier phase difference,
$\phi_{31} = \phi_3 - 3\phi_1$.  \citet{sandage04} scaled the
metallicities obtained from the \citet{jk96} period-phase-\feh relation 
to the \citet{ziwe} system, and we adopt his scaling relation:

\begin{equation}
\label{makezw}
{\rm \feh}=-5.490-5.664P+1.412\phi_{31}.
\end{equation}
The metallicity scale adopted throughout this paper is on the 
\citet{ziwe} scale.  

\begin{figure}[htb]
\includegraphics[width=16cm]{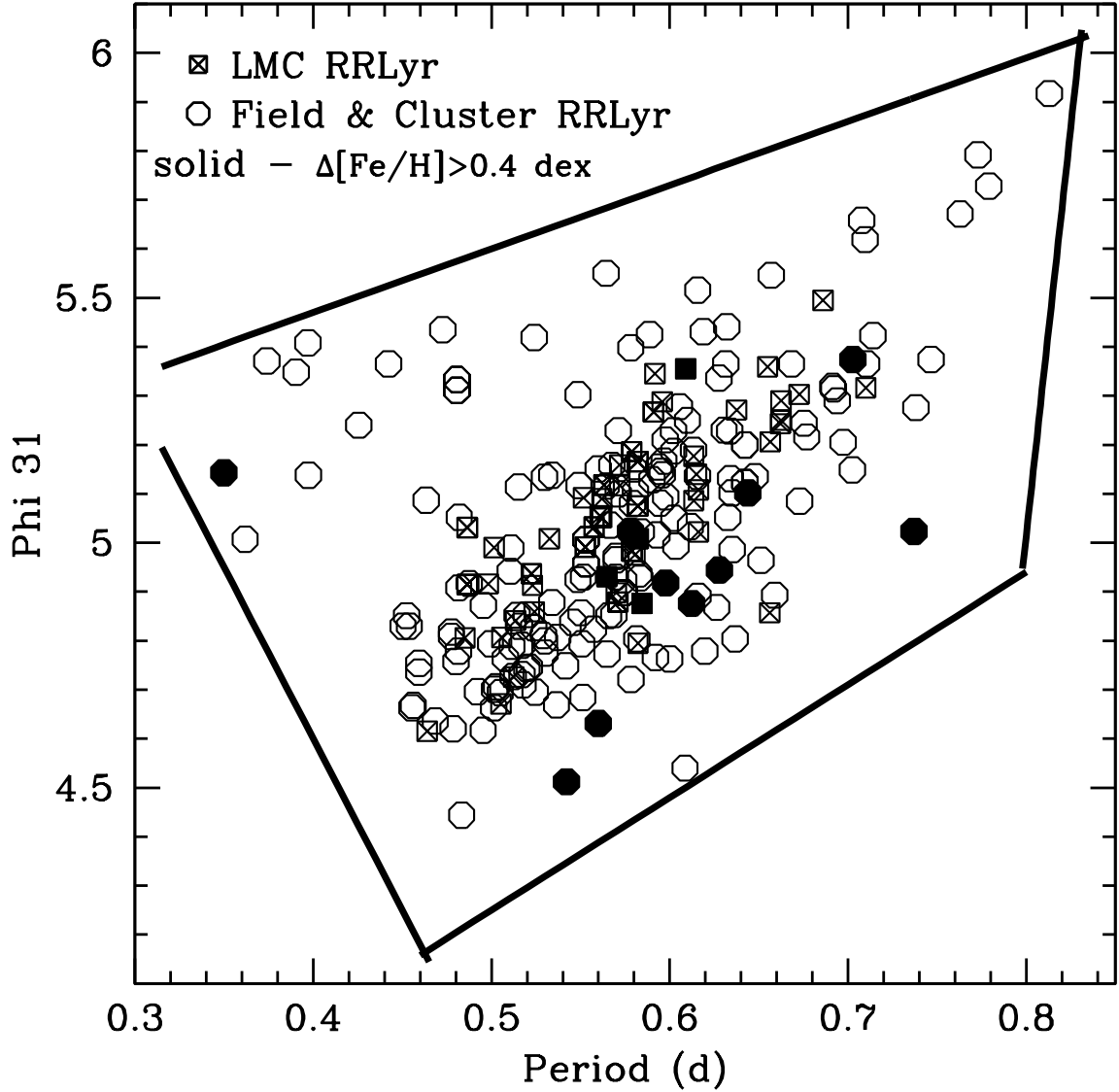}
\caption{A period-$\phi_{31}$ diagram of the calibrating RR0 Lyrae
sample.
The squares indicate the 53 MACHO LMC RR0 Lyrae stars, 
the circles represent the 177 field and cluster
RR0 Lyrae stars.  The filled symbols represent those RR0
Lyrae stars that have photometric metallicities derived from 
Equation~\ref{makezw} that deviate by more than 0.4 dex from their
spectroscopic metallicities.  The enclosed region indicates the area
in the period-$\phi_{31}$ plane in which Equation~\ref{makezw} is
taken to be valid.
\label{plottwo}}
\end{figure}

Figure~\ref{plottwo} shows the location of our calibration stars in the 
$\rm period-\phi_{31}$ plane.  These stars include 61 RR0 Lyrae field stars 
from the \citet{jk96} sample, 
116 RR0 Lyrae stars from the clusters Omega Cen, M68, IC 4499, NGC 6171, 
M3 and M92 and 53 MACHO LMC RR0 Lyrae stars, all which have
spectroscopic \feh measurements.  These calibration stars are discussed in 
more detail in the Appendix.
The solid lines in Figure~\ref{plottwo} enclose the area in which
the calibration stars fall and
is the region in which we take Equation~\ref{makezw} to be valid.  In
this paper, we will exclusively use the RR0 Lyrae stars that fall within
this region.
The completeness implications of this period-$\rm \phi_{31}$ criterion is 
discussed in the \S 4.

The metallicity catalog for the 2523 bulge stars is presented in 
Table~\ref{tab1};
the 255 Sgr stars are presented in Table~\ref{tab2}.  
The $V$ and $R$ magnitudes given are the average magnitudes.  
As shown in the Appendix, based upon MACHO LMC
photometric metallicities, the error in the \feh values is $\pm$0.2 dex. 

\subsection{The deviation parameter selection effect}
Here we examine what stars are most affected by
implementing the $D_m$ criteria and how that would affect statistical
properties of the sample.  Large values of the deviation parameter can be 
due to RR Lyrae which have
intrinsically unusual lightcurves, or simply poorly sampled and/or
noisy observational data.  Especially the later effect could introduce a bias
when determining the average metallicity of the sample.

We first look for a spatial bias by binning the RR0 Lyrae stars by 
Galactic coordinates.  We define a completion ratio ($f$), as the fraction 
of stars with a $D_m <$ 4.5.  A careful inspection of the resulting 
completeness map found no trend in the completion-ratio with
location in the sky.

\begin{figure}[htb]
  \includegraphics[width=16cm]{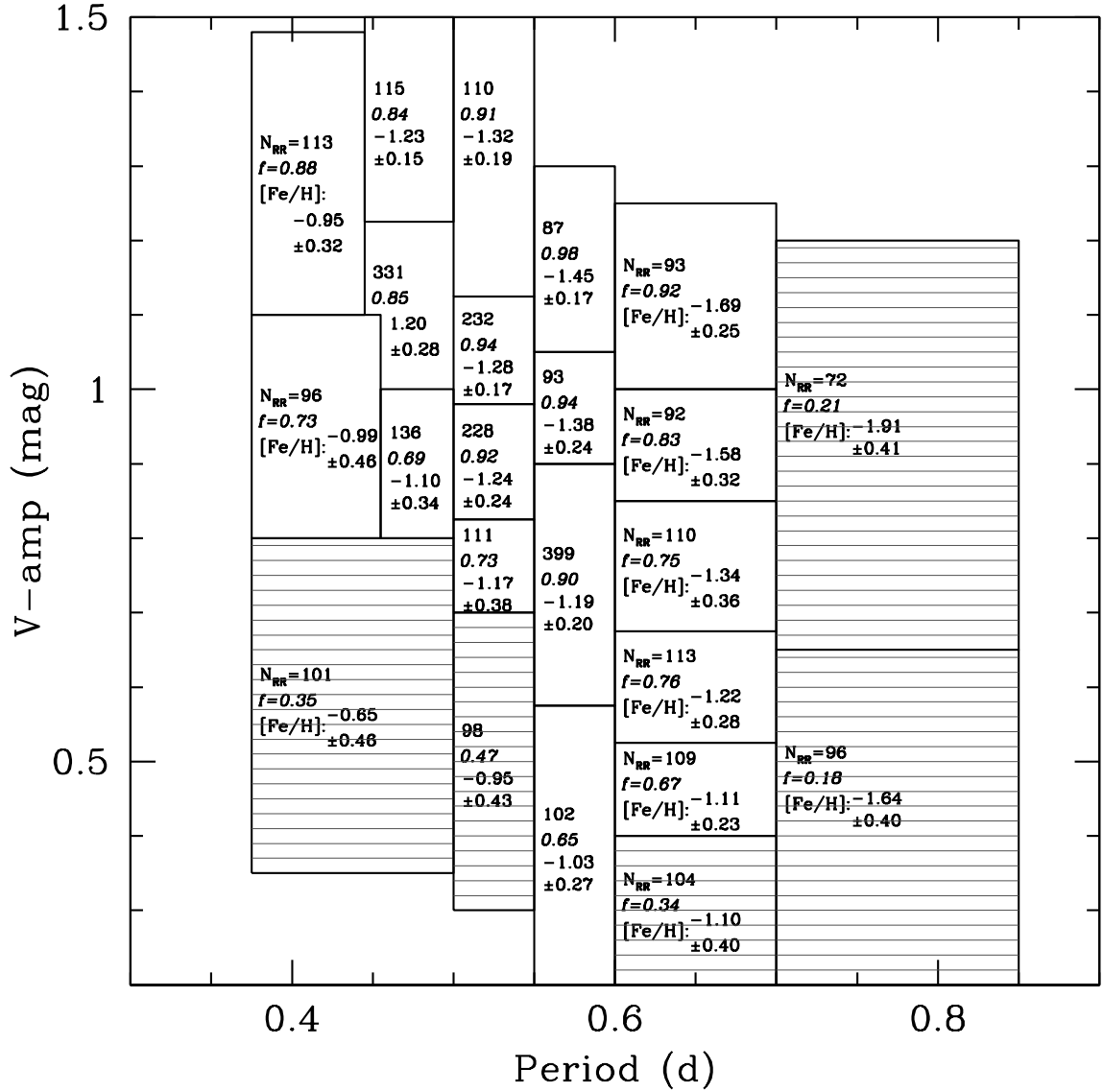}
  \caption{A period-$V$-amplitude diagram of all 3260 MACHO bulge RR0 
Lyrae stars binned in $\sim$ 100 RR0 Lyrae bins.  The total 
number of RR0 Lyrae in each bin, $N_{RR}$, the statistic, $f$, 
defined as the fraction of stars with a $D_m >$ 4.5, and
the average $\rm \feh$ and dispersion (as calculated from 
Formula~\ref{makezw}) is shown in each bin.  Bins with a 
similar $f$-ratio,  $\rm \feh$ and dispersion are combined.
The shaded boxes illustrate where $f$ $\rm < 60\%$ lie.  
\label{plotthree}}
\end{figure}

Figure~\ref{plotthree} is a Period-$V$-amplitude (PA) diagram that encompasses 
the range of all the 3260 MACHO bulge RR0 Lyrae stars.  The RR0 Lyrae 
stars are binned in the
Period-$V$-amplitude plane so that each bin has $\sim$ 100 RR0 Lyraes, and
the bins are represented by boxes.  

The shaded boxes illustrate regions where $f$ $\rm < 60\%$, and
hence where RR0 Lyrae stars are most affected by the $D_m$ criteria.
These bins have $\sim$ 450 
stars in them, or $\sim$ 14 \% of the sample.  The bins 
where $f$ $\rm < 75\%$ comprise 30 \%  
of the sample.

Stars with short periods and small amplitudes 
(P$ <$ 0.5 d, $V$-amp $<$ 0.8 mags), as well as the stars with long periods 
(P $>$ 0.7 d) are likely to have too high of a deviation parameter to be 
included in a metallicity determination.  
There are three bins in the bulge sample with $f$-ratios of 
0.35, 0.46, and 0.34 that have an average photometric metallicity from
Equation~\ref{makezw} that are more metal rich than the average metallicity of 
$\rm <\feh> \sim -$1.3 dex.  If all the stars in these bins that 
currently have
$D_m >$ 4.5 (and hence are not included in our metallicity analysis)
are given a \feh equal to that of the average metallicity of their 
respective bin, then the average metallicity of the bulge RR0 Lyrae sample 
would increase by $\sim$ 0.025 dex and the dispersion would increase by
$\sim$ 0.004.

Similarly, there are two bins in the bulge sample with $f$-ratios of
0.18 and 0.21 that have average photometric metallicities more
metal poor than \feh = $-$1.3 dex.  
The stars in these bins are $\sim$ 5\% of our sample, and 
the bias from either an incorrect or no metallicity estimate is again
small.  By and large, the completion ratio is large over the PA-diagram.  
There may be some incompletion effects with the long period stars, as well 
as the small amplitude/short period stars, but there are few stars in 
these bins, and the mean properties of the sample won't change significantly.

\section{Analysis of \feh results}
\begin{figure}[htb]
  \includegraphics[width=16cm]{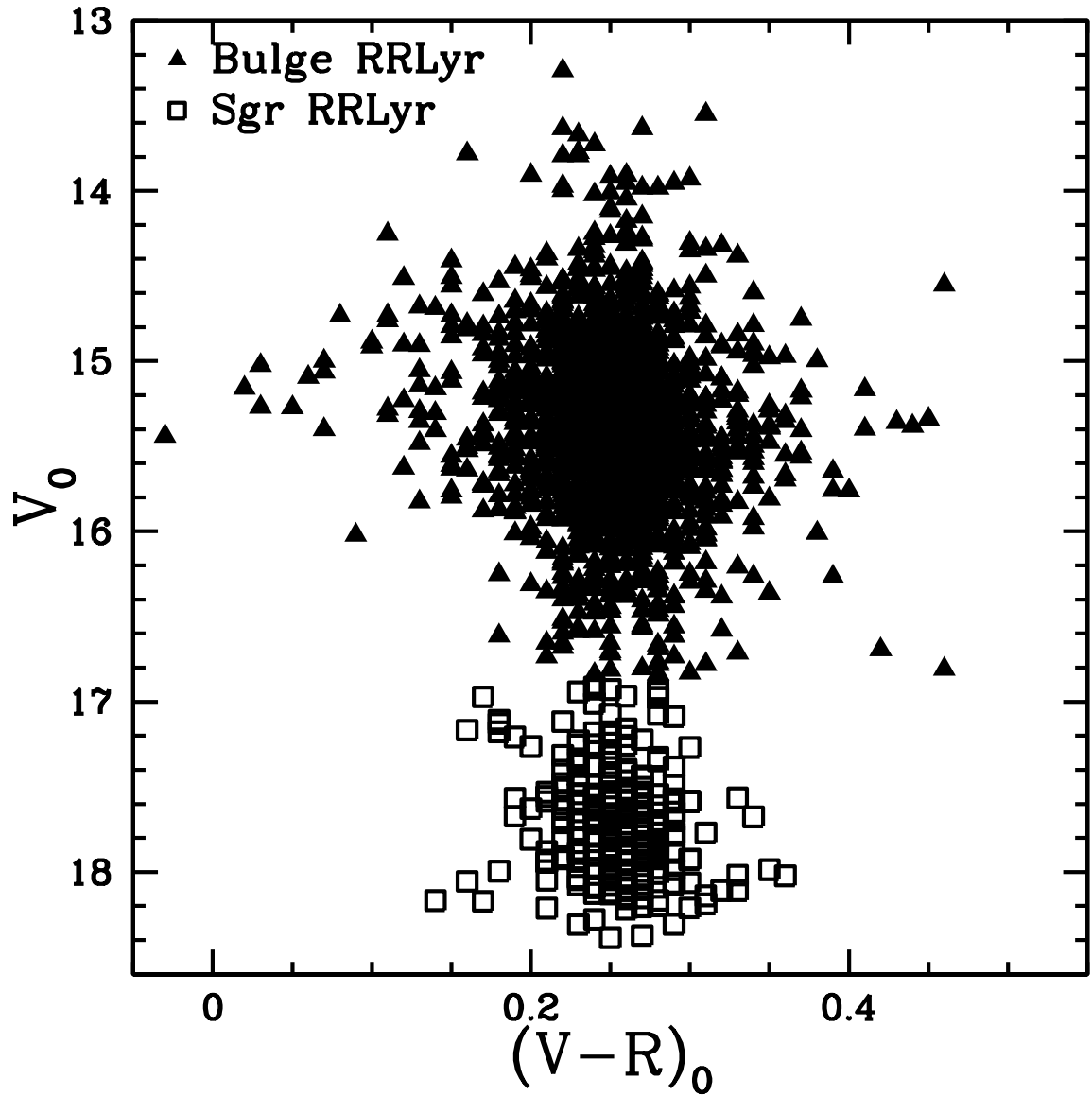}
  \caption{A dereddened color magnitude diagram of the RR0 Lyrae
stars from the MACHO bulge fields with a $D_m < 4.5$.
\label{plotfour}}
\end{figure}

There are 2690 RR0 Lyrae stars with a $D_m < 4.5$ that fall within
the period-$\phi_{31}$ calibration region.  Most of the RR Lyrae
stars have magnitudes that peak at $V =$ 15, which places them at
a distance of 8 kpc.  A few, however, are more than twice as far away, 
and these
distant stars belong to the neighboring Sagittarius dwarf galaxy (Sgr).  
In order to separate these two populations, the RR0 Lyrae stars are 
dereddened using the \citet{kunder08} reddenings, and the dereddened
color-magnitude diagram is shown in Figure~\ref{plotfour}. 
The two populations are separated at $V_{0} =$ 16.9, where the
smaller (brighter) $V_{0}$ magnitudes represent the bulge RR Lyrae stars; 
the fainter RR Lyrae stars belong to the Sagittarius dwarf galaxy.  
The $V_{0} =$ 16.9 division is somewhat arbitrary, but changing the dividing
line has no major impact on the conclusions drawn here.  
Following this procedure, 2435 RR0
Lyrae stars are assigned to the bulge and 255 RR0 Lyrae stars are
assigned to Sgr dwarf galaxy.

\subsection{Galactic Bulge RR Lyrae Stars}
\begin{figure}[htb]
  \includegraphics[width=16cm]{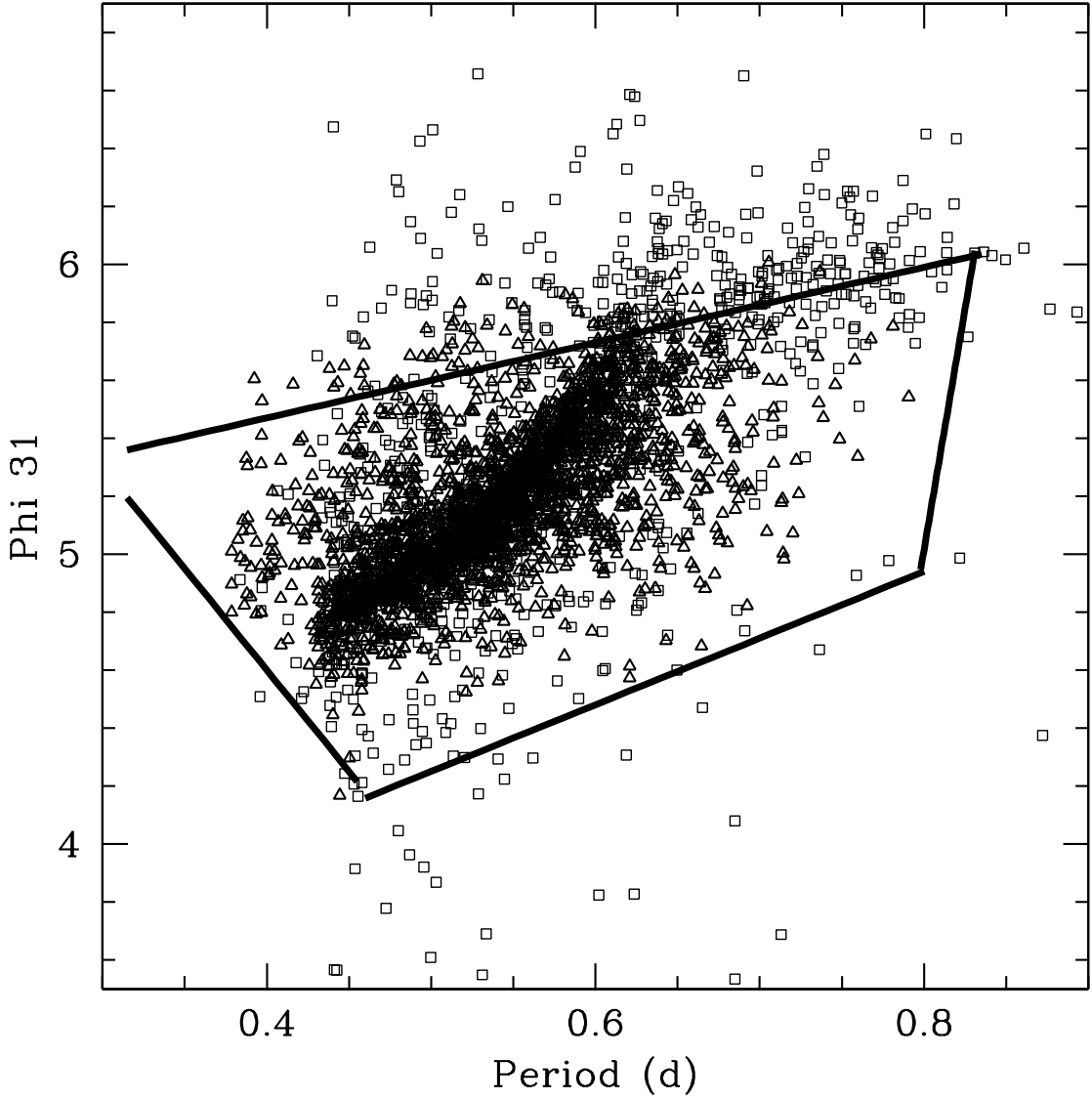}
  \caption{A period-$\phi_{31}$ diagram of all 3260 MACHO Bulge RR0 
Lyrae stars.  The triangles represent stars with a $D_m < 4.5$, and
the squares indicate which stars do not have a favorable $D_m$.
 There are 360 RR0 Lyrae stars that fall outside the 
``calibration region'' and 80 of those have a $D_m$ that
would allow for a \feh determination.
\label{plotfive}}
\end{figure}

Figure~\ref{plotfive} shows the location of MACHO Bulge RR0 Lyrae stars
in the $\rm period-\phi_{31}$ plane with the same enclosed region 
as discussed in \S 3 over-plotted.
There are 360 Bulge RR0 Lyrae stars, or 15\% of the bulge sample, 
which fall outside this area; only 80 of those (3 \%)
have a $D_m < $ 4.5.  The $f$-ratio, or fraction 
of stars with a $D_m <$ 4.5, significantly decreases outside 
the calibration region.   Thus, the average \feh is not significantly affected 
by the $\rm period-\phi_{31}$ criterion.

\begin{figure}[htb]
\includegraphics[width=16cm]{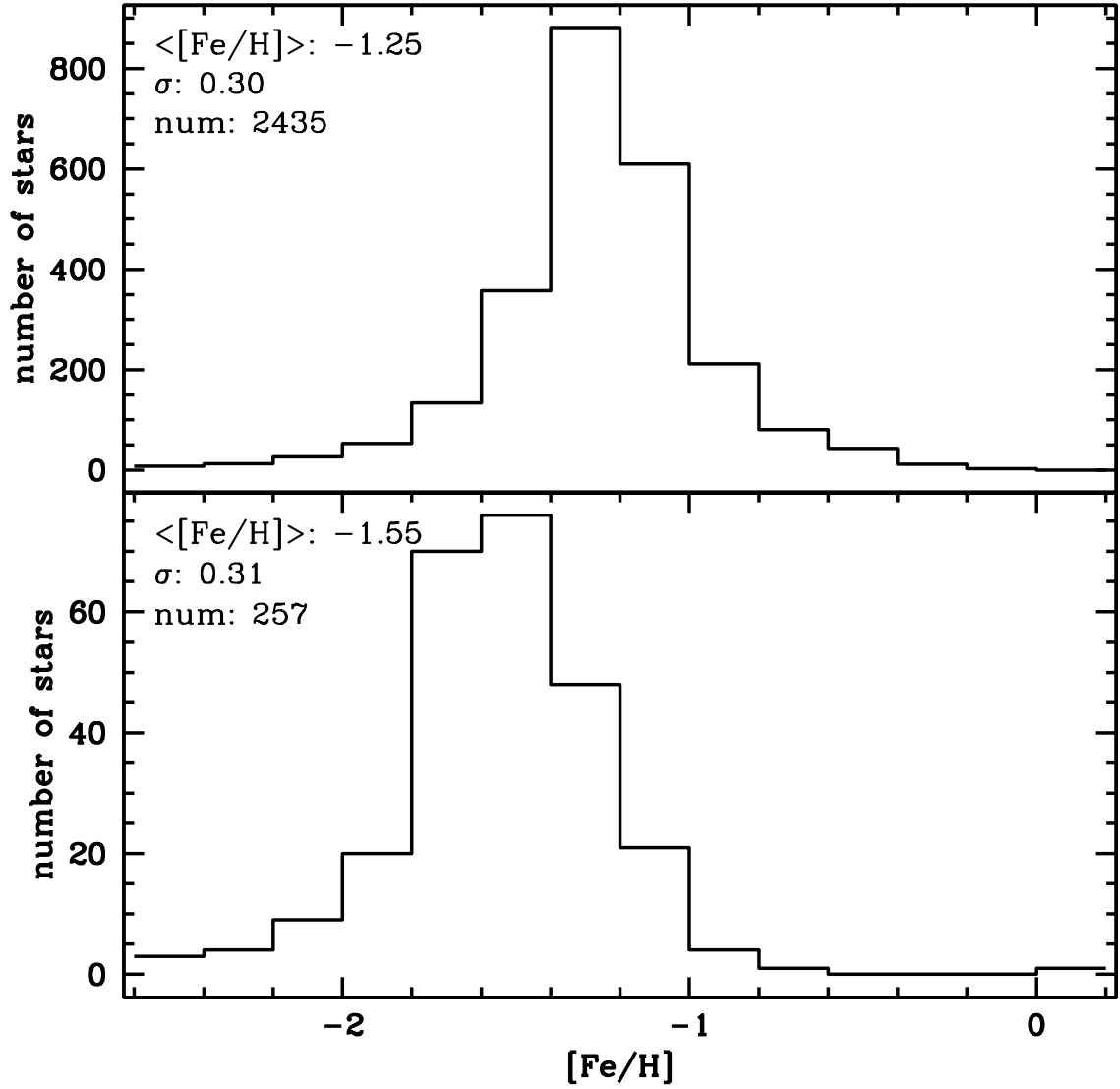}
\caption{\feh of RR0 Lyrae stars in the Galactic Bulge (top)
and Sgr (bottom) obtained using equation~\ref{makezw}.  The number
of stars, $<{\rm \feh}>$, and the standard deviation, $\sigma$, 
is given in the top left corner of each figure.
\label{plotsix}}
\end{figure}

The top histogram in Figure~\ref{plotsix} shows
the metallicities of the 2435 Galactic Bulge RR0 Lyrae stars.  
We find that RR0 Lyrae stars in the bulge have $<{\rm \feh}>$ $=
-1.25 \pm 0.01$.  This average is
consistent with the spectroscopic one \citep{walkThisTurn},
and is also near the lower limit for the
distribution of K giant abundances in the same field 
(Rich 1998; Geisler \& Friel 1990).
The dispersion found from the Galactic Bulge RR0 Lyrae stars is 0.30 dex.  
Using our metallicity error of
$0.20$ dex, we obtain the intrinsic metallicity dispersion of 
the bulge RR0 stars of 0.22 dex.  Note that the dispersion of 
the RR Lyrae variables is larger than expected from
the measuring errors alone, which is evidence for a real abundance
range.  Due to the nature of horizontal 
branch evolution, RR Lyrae stars preferentially come from metal 
poor populations, and so the \feh may be biased
compared to the general population of old stars in the bulge.  

The central 99\% of the RR0 Lyrae metallicities range from 
\feh$= -2.26$ to $-$0.15 dex.  This is a wider spread 
than that found by \citet{walkThisTurn}, who took spectra of
59 RR Lyrae variables in BW.  However, our sample is 
40 times greater and covers a larger area.  
Our results are in reasonable agreement with \citet{blanc84}, 
who determined abundances 
indirectly for 51 RR0 variables in BW by 
extrapolating from the positions of the M3 and M15 variables in the
period-amplitude diagram  (Sandage, Katem, \& Sandage 1981; 
Sandage 1982) and found a distribution ranging from \feh $= -2.2$ to 
$+0.4$ dex.  Our bulge metallicity distribution is also consistent 
with \citet{zoccali}, who analyzed near IR photometry of 513 bulge 
giants and inferred metallicities from \feh $= -2.0$ to +0.7 dex.   

\begin{figure}[htb]
\includegraphics[width=16cm]{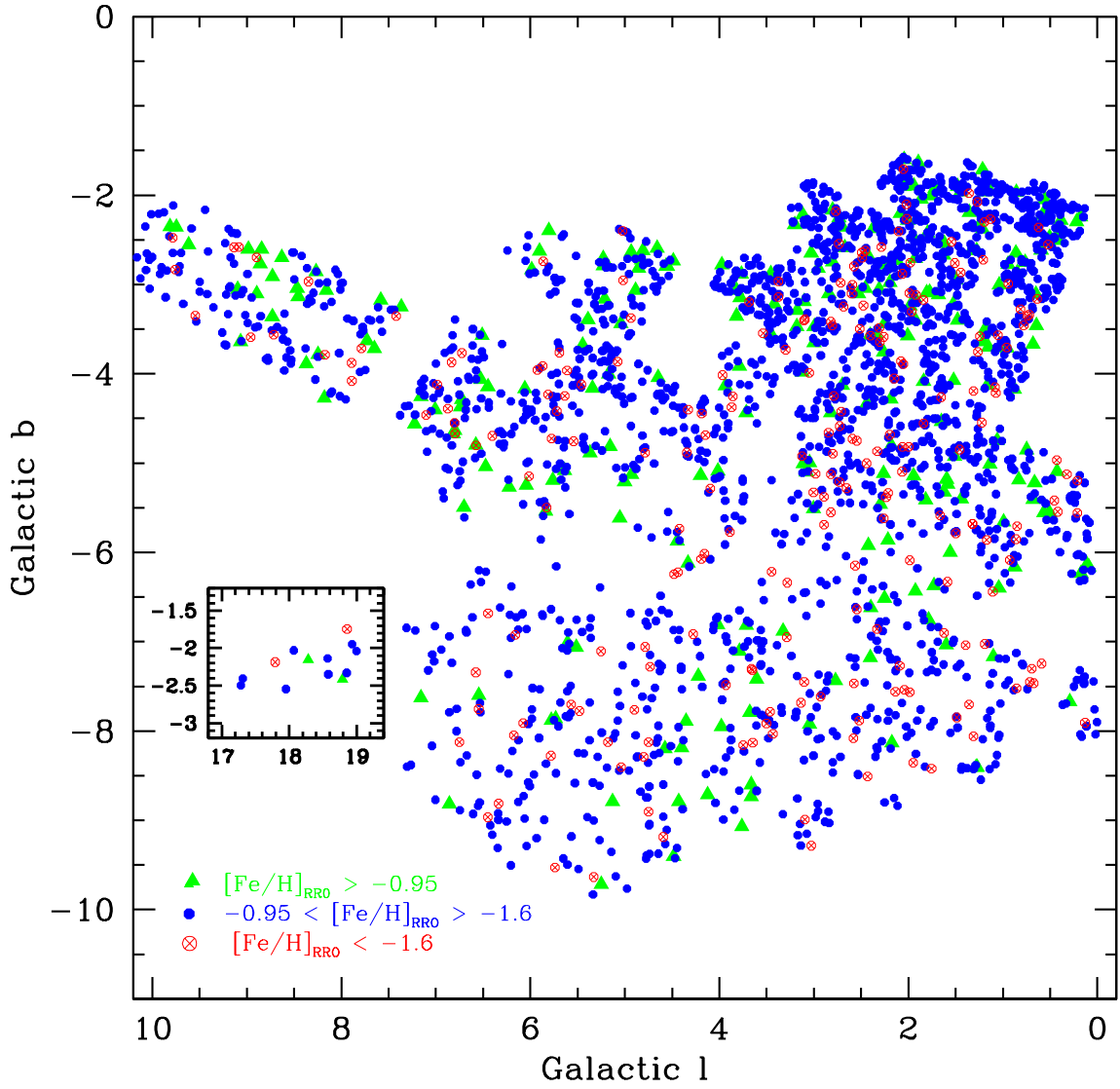}
\caption{Metallicities of MACHO RR0 Lyrae stars as a function of Galactic
latitude and longitude.  The insert presents three disk fields 
separated from the others by 10 $^\circ$.
\label{plotseven}}
\end{figure}

The bulge RR0 stars are separated into three metallicity bins 
and plotted according to their position in the sky in Figure~\ref{plotseven}. 
It is suggestive from Figure~\ref{plotseven} 
that the fields with $l<4^{\circ}$ and $|b|<3.5^{\circ}$ have 83$\pm$3\% 
RR0 Lyrae stars with ${\rm \feh}$ between $-0.95$ and $-1.6$ dex.  
The fields outside this range consist of RR0 Lyrae stars of which
78$\pm$2\% have an intermediate metallicity.  This is slight indication 
that RR0 Lyrae
stars further from the central bulge ($l<4^{\circ}$ and $|b|<3.5^{\circ}$), 
seem to vary more in their metallicities. 

It is an interesting question how our metallicities compare not just on 
average but on a star-by-star basis.
\citet{walkThisTurn} took low-resolution spectra of 42 RR0 Lyrae stars in BW, 
and via the $\Delta S$ method derived metallicities for these stars.  
Based on position in the sky and
period, 32 MACHO RR0 stars could be matched with the ones from their sample.
Five of these had deviation parameters too large for an accurate metallicity
derivation; this leaves us with 27 \citet{walkThisTurn}
RR0 stars available for a comparison.  
\citet{rodgers77} find the $\rm [Ca/H]$ of 27 stars in the Palomar-Groningen
fields 2 and 3 of the galactic nuclear bulge.  Two of the stars in this
sample overlap with stars in the MACHO bulge sample.  Using \citet{rodgers74}
dependence of $\rm [Ca/H]$ on $\Delta S$, (their Table~2), a $\Delta S$
can be assigned to each of these stars, and their $\rm \feh$ can be
found.\footnote{The same  $\Delta S$ - $\rm \feh$ relation used in
\citet{walkThisTurn} is employed here.  This relation is from
\citet{butler75} on the \citet{ziwe} metallicity scale.}

\begin{figure}[htb]
\includegraphics[width=16cm]{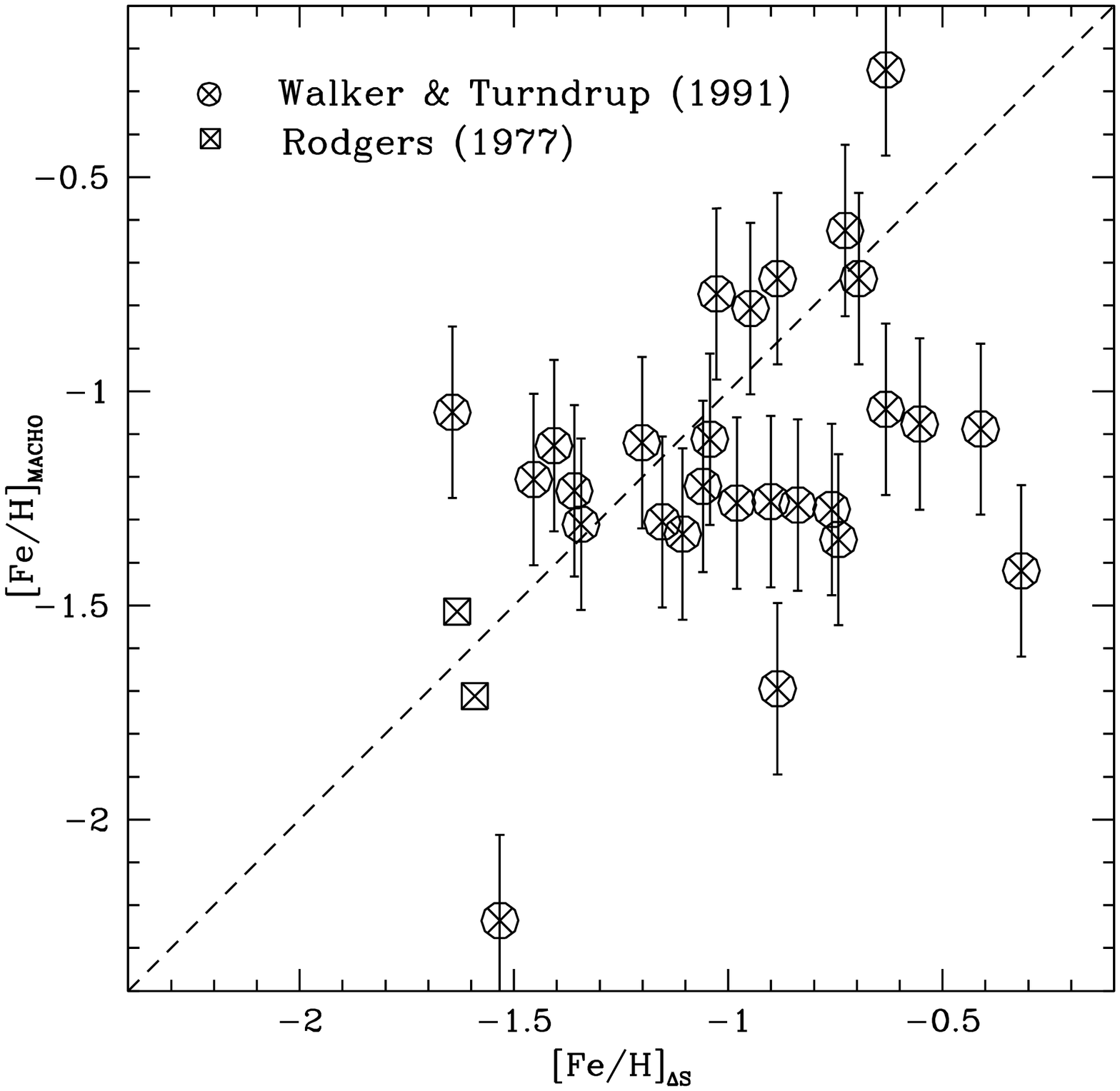}
\caption{Comparison between the metallicity from
previous $\Delta S$ observations, ${\rm \feh_{\Delta S}}$ 
\citep{walkThisTurn}, 
and that from our Fourier coefficients, ${\rm \feh_{MACHO}}$.
\label{ploteight}}
\end{figure}

Figure~\ref{ploteight} shows the comparison between the metallicity
based on  $\Delta S$ observations \citep{walkThisTurn, rodgers77}
and that from our Fourier coefficients.
There are some large differences between the \feh determined 
here, and \feh determined from the $\Delta$S measurements of 
\citet{walkThisTurn}.  In particular, we determine a smaller range of 
abundances than that found by \citet{walkThisTurn}.  
The dispersion about the line of unity is 0.42 dex.
Unlike spectroscopic methods that rely upon measurements of the 
CaII K-line, the Fourier method does not require a knowledge
of the interstellar K-line strength. Those lines can be strong enough to
sometimes be a significant source of uncertainty in the application of the
$\Delta$S method to bulge RR Lyrae stars.
The agreement between the MACHO LMC photometric \feh determinations and 
the LMC \feh spectroscopy from \citet{bori06} and \citet{grat04} 
as discussed in the Appendix, 
provides evidence that the photometric metallicities here are reliable.  
More spectra of RR0 Lyrae stars in 
the bulge, especially intermediate or high-resolution spectra, is needed 
to clarify this discrepancy.

\subsubsection{Presence of two different populations}


The elongation of the bulge and its barred structure was clearly shown by
the DIRBE IR maps \citep{dwek95}, and by the clump giants \citep{stan94}.
\citet{cabrera07} use the red-clump population to show that two very different 
large-scale triaxial structures coexist in the inner Galaxy, a
thick bulge and a long thin bar.  The long thin bar is constrained
to  $|b| < 2.0^{\circ}$, whereas the thick bulge extends to at least
$|b| < 7.5^{\circ}$.  Although RR Lyrae stars are classical tracers 
of old metal-poor populations, a RR Lyrae barred distribution is 
not as clear as that of bulge clump giants.  For example, \citet{alc98} 
found evidence of a bar using red clump 
stars in the MACHO fields, but showed that the bulk of RR Lyrae stars 
do not follow the expected barred distribution.
A small barred structure may be present in the bulge RR Lyrae population
(Collinge, Sumi \& Fabrycky 2006), but other studies 
\citep{alard96, wesselink87} find no bar in the RR Lyrae distribution.
In fact, as \citet{alc98} point out, the RR Lyrae stars analyzed
by \citet{wesselink87} tended to be fainter and, therefore, more distant
than those at negative longitudes, when forcing a triaxial fit to his 
low-latitude data.  

The absence of a strong bar in the RR Lyrae population
would indicate that they represent a different population than the
metal rich bulge.  
Using the effect of metallicity in the period-amplitude plane, 
\citet{alc98} separated
RR Lyrae stars into three bins containing the metal-rich, intermediate,
and metal-poor stars.  Only for the metal-rich subsample could a barred 
distribution not be ruled out within their errors.

We look at the mean RR0 Lyrae magnitudes falling within 
$|b| < 3.5^{\circ}$.  
For a barred distribution with a standard inclination angle,
one would expect that RR Lyrae stars at positive Galactic longitudes
would be brighter since they would be nearer than those at negative
longitudes.  Reddening-independent magnitudes are used to avoid
uncertain reddening corrections and are defined as $W_V = V - 4.3(V-R)$, 
where the factor
4.3 is the selective extinction coefficient $R_{V,VR}$ derived
by \citet{kunder08}.  The dominant error in the MACHO photometry
comes from zero-point uncertainties between different MACHO fields.
For example,  
Cook et al. \citetext{in preparation} estimate a $\sim 0.16$ error in the $V$-magnitudes 
from overlap RR0 Lyrae stars in the MACHO fields.
To remove systematic magnitude errors that occur between fields, the RR Lyrae 
magnitudes are binned by MACHO field; the statistical 
error associated with each $W_V$ magnitude bin is 0.16 mags. 

\begin{figure}[htb]
\includegraphics[width=16cm]{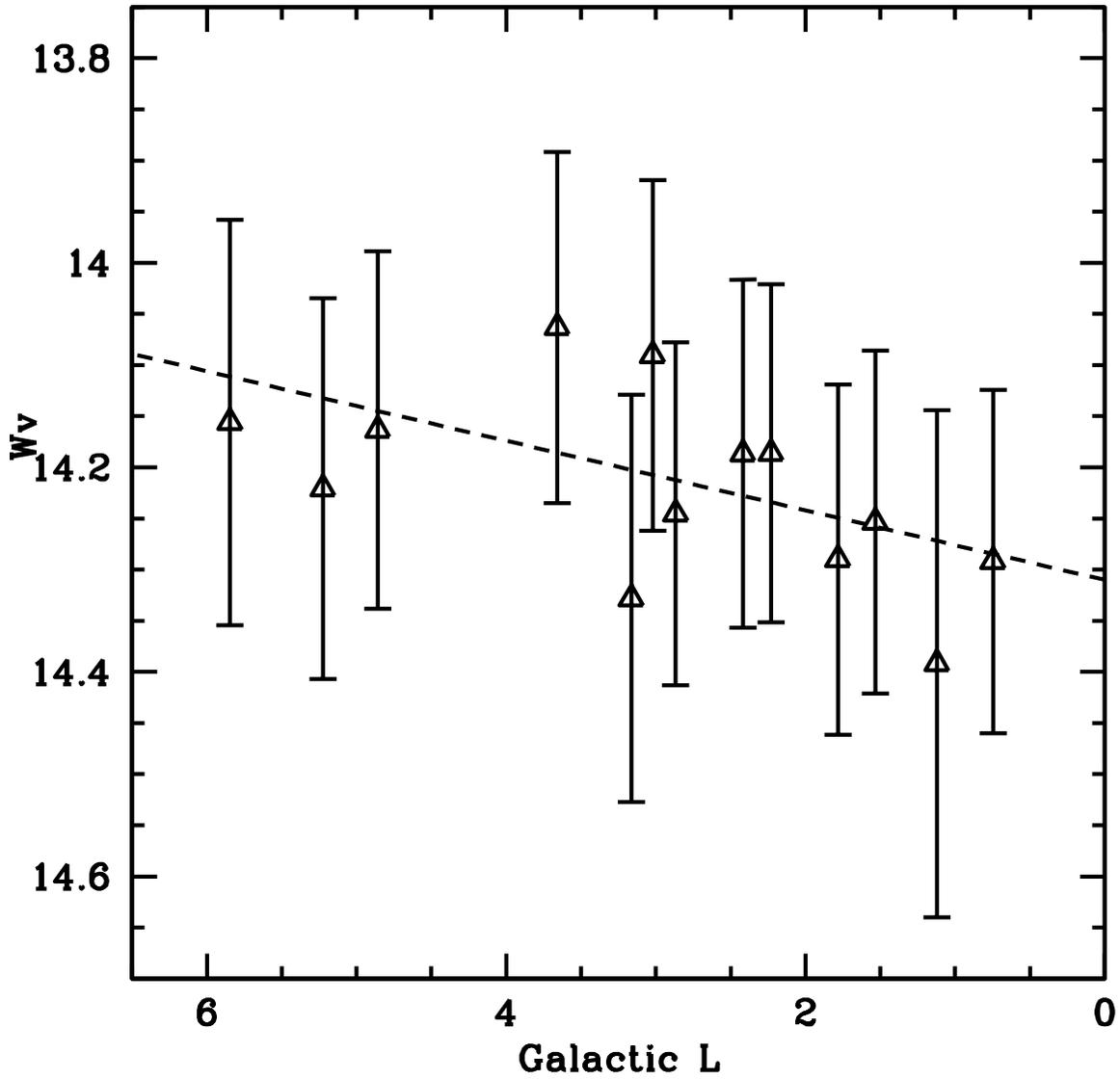}
\caption{Mean magnitudes of RR0 stars with $b > -3.5$ versus 
Galactic longitude.  The dashed line is a least squares fit to the
data, which has a slope of ($-0.034 \pm$0.015) $\rm mag/^{\circ}$.
\label{plotnine}}
\end{figure}

Figure~\ref{plotnine} shows the mean
reddening-independent magnitudes in each of the MACHO bulge fields
with $|b| < 3.5^{\circ}$.  In each field, there are between 4 and 100 
RR0 Lyrae stars.  Assuming no positional error in each bin, the 
weighted $<W_V>$ on galactic L regressions to all the stars shows that the 
RR0 Lyrae stars show a slight signature of the expected barred distribution
with a slope of ($-0.034 \pm$0.015)$\rm mag/^{\circ}$.  Restricting the 
sample to
include only those stars with $l < 3^{\circ}$, the barred signature
becomes stronger with a slope of $-0.07 \pm 0.03$.  Both of these values
are in agreement with the barred signature found in the Red Clump stars 
\citep[0.037 mag/$^\circ$][]{stan94} and in the OGLE Bulge RR0 Lyrae 
stars \citep[0.038 mag/$^\circ$][]{collinge06}.


Given the uncertainties in the MACHO $V$-band photometry, the RR Lyrae 
absolute magnitude dispersion, and the depth of the bulge, a simple 
Monte Carlo analysis is performed to see if a
bar could in fact be seen in the MACHO data.  
A Gaussian dispersion about 14.5 mags is generated for each MACHO field
that takes into account an RR Lyrae absolute
magnitude dispersion of 0.1 mags, a bulge distance spread
of 0.29 mags (which corresponds to a 1 kpc distance spread), 
and the 0.16 mag photometric error.  Next a slope of 
$\rm -0.037$ $\rm mag/^{\circ}$ is assigned to this data.
We find that the given slope is recovered from the simulated $W_V$ values
with a 5-sigma detection.  The 2-sigma barred detection found in the MACHO
data is indication that the RR Lyrae stars only weakly trace out the bar.

The MACHO Bulge RR0 Lyrae stars were separated into metal-rich,
intermediate, and metal-poor metallicity bins and the procedure is repeated.  
This reduces the amount of RR Lyrae stars in each field, however, and the 
Monte Carlo analysis predicts that no conclusive bar-signature would be 
seen.  Indeed a quantitative analysis of the RR0 Lyrae stars in these
bins reveals no slope for either populations.

Although a barred slope potentially could be seen with the MACHO data,
no strong bar signature is seen with the RR0 Lyrae stars here.  This is further
evidence that the majority of the RR Lyrae stars do not belong to the 
triaxial bulge, 
but belong to the extension of the Galactic halo in the inner regions.
The absence of a strong bar in even the RR Lyrae population needs to be 
considered when forming scenarios on the formation of the Milky Way bulge.

\subsection{RR Lyrae Stars from Sagittarius dwarf galaxy}
\begin{figure}[htb]
\includegraphics[width=16cm]{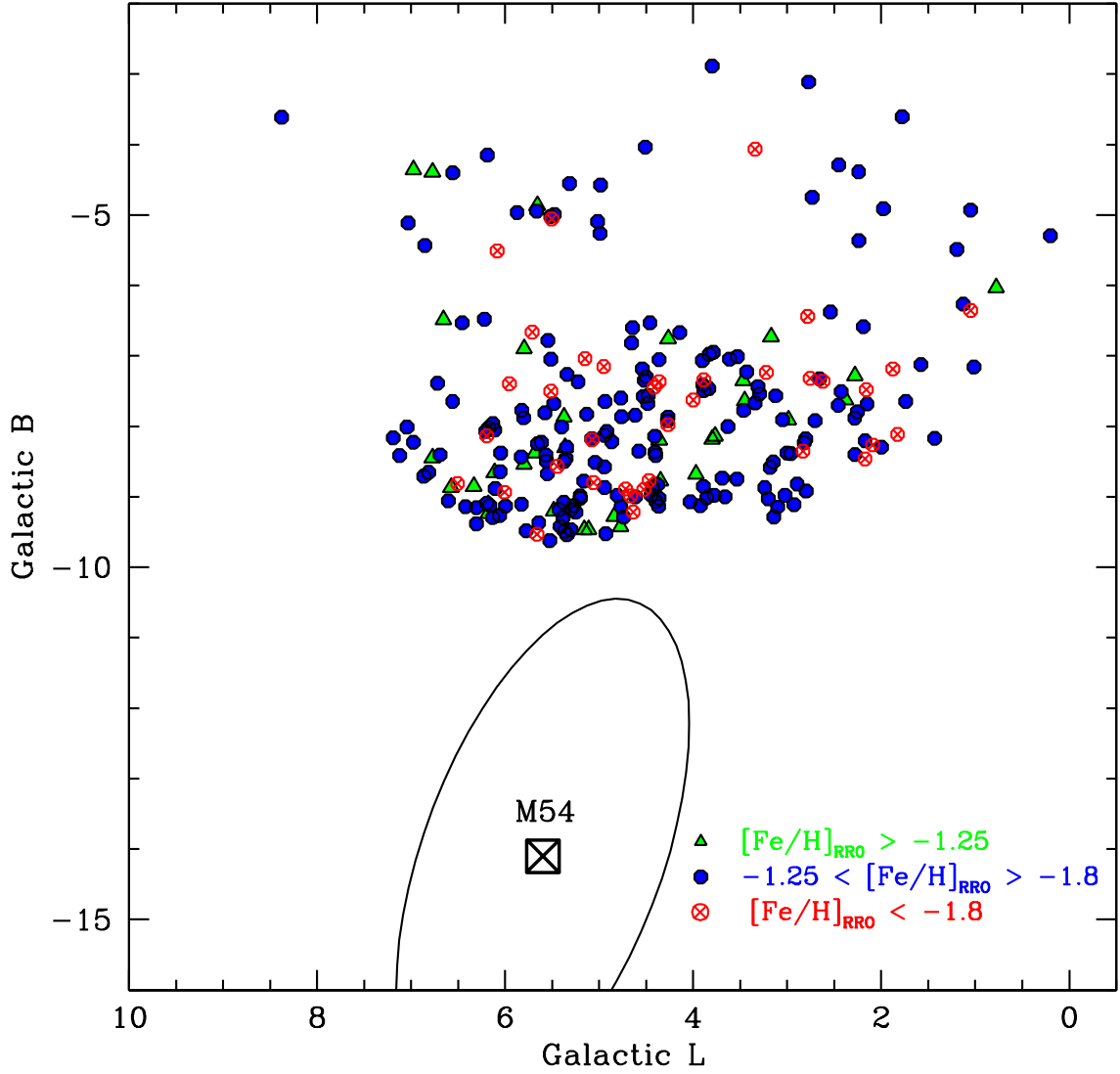}
\caption{Metallicities of the MACHO Sgr RR0 Lyrae stars as a function of 
Galactic latitude and longitude.  The location of M54 is displayed
as well as the core radius of the Sgr galaxy as traced out from M giants 
\citep[assuming an ellipticity of 0.65 and a position angle of 
104$^\circ$;][]{majewski03}.
\label{plotten}}
\end{figure}

The Sagittarius dwarf spheroidal galaxy 
(Sgr) is a nearby satellite of the Milky Way and one of the galaxies
that is currently disrupting under the strain of the Galactic tidal field.
(Ibata et al.\ 1994, 1997, Monaco et al.\ 2004, Majewski et al.\ 2003).  It
spans a large area and contains a mix of stellar populations.  
Sixteen Sgr RR0 Lyrae stars, or 6\% of the Sgr sample, fall outside the 
 $\rm period-\phi_{31}$ criterion shown in Figure~\ref{plottwo} and discussed
in \S3.  As none of these stars have a favorable $D_m$, the average
\feh is not affected.  Figure~\ref{plotten} shows the location of the 
Sgr stars in the MACHO sample relative to the center of the galaxy, 
assumed to be M54.  The RR0 Lyrae stars are separated according to
their metallicity in this figure, and no trend of metallicity with 
location in the sky is apparent.  

The mean metallicity of the RR0 stars in area of the Sgr covered by the MACHO 
survey is $<{\rm \feh}> -1.55 \pm 0.02$.    
This is based on 255 RR0 Lyrae stars with a dispersion of 0.31.  The
intrinsic SGR metallicity dispersion is 0.25 dex, 
identical to that in the bulge, and identical to the
internal dispersion found by \citet{ald01} of giant branch stars.  
The metallicity range for the central 99\% stars is $\rm \feh = -2.41$ to 
$-$0.76.  These results are consistent with \citet{ald01} and 
\citet{nature}.  It is
also in agreement with \citet{mateo}, who used OGLE RR Lyrae stars to 
find that the properties of the Sgr giant branch and upper main 
sequence is consistent with a mean metallicity of 
${\rm \feh} \sim -1.2 \pm 0.3$.  

\section{Distances to the Stars}
RR Lyrae stars are well known standard candles and play an important
step in the distance ladder.  It has long been known that their
absolute magnitude is a function of metallicity, and it has
normally been assumed that the absolute magnitude of an 
RR Lyrae star has a linear dependence on \feh \citep[e.g.][]{krauChab}.  
In recent work, it has been suggested that this linear dependence
is not suitable for the most metal-rich (\feh $> -$0.7 dex)
field variables (Bono, Caputo, \& di Criscienzo 2007).
We use the quadratic relation from \citet{bono07}:
\begin{equation}
\label{distmv}
M_V = 1.19 + 0.5\feh + 0.09\feh^2
\end{equation}
to find the absolute magnitude of each star.  This is valid over the entire
metallicity range of $\rm \feh = -2.5$ to $\sim$ 0.

Using Equation~\ref{distmv} the mode of the Bulge RR0 Lyrae star sample 
is 8.02 kpc $\pm$ 0.05 where the error indicates the uncertainty in 
determining the mode only.  Taking a reasonable error in the RR Lyrae
absolute magnitude-\feh zero-point calibration (0.08 mags) and propagating it
linearly with the MACHO zero-point and calibration error,
the estimated error in the distance to the Galactic Bulge is 0.3 kpc.
Therefore, our estimated distance to the Galactic Bulge is
8.0 $\pm$ 0.3 kpc.
This is in reasonable agreement with \citet{eisenhauer05}, who found 
$R_0 =$ 7.62 $\pm$ 0.32 kpc from the orbit of one star around
the Milky Way's central black hole.  It is also in agreement with 
Groenewegen, Udalski \& Bono (2008) who found $R_0 =$ 7.94 $\pm$ 0.26 kpc
from K-band magnitudes of Population-II Cepheids. 

\begin{figure}[htb]
\includegraphics[width=16cm]{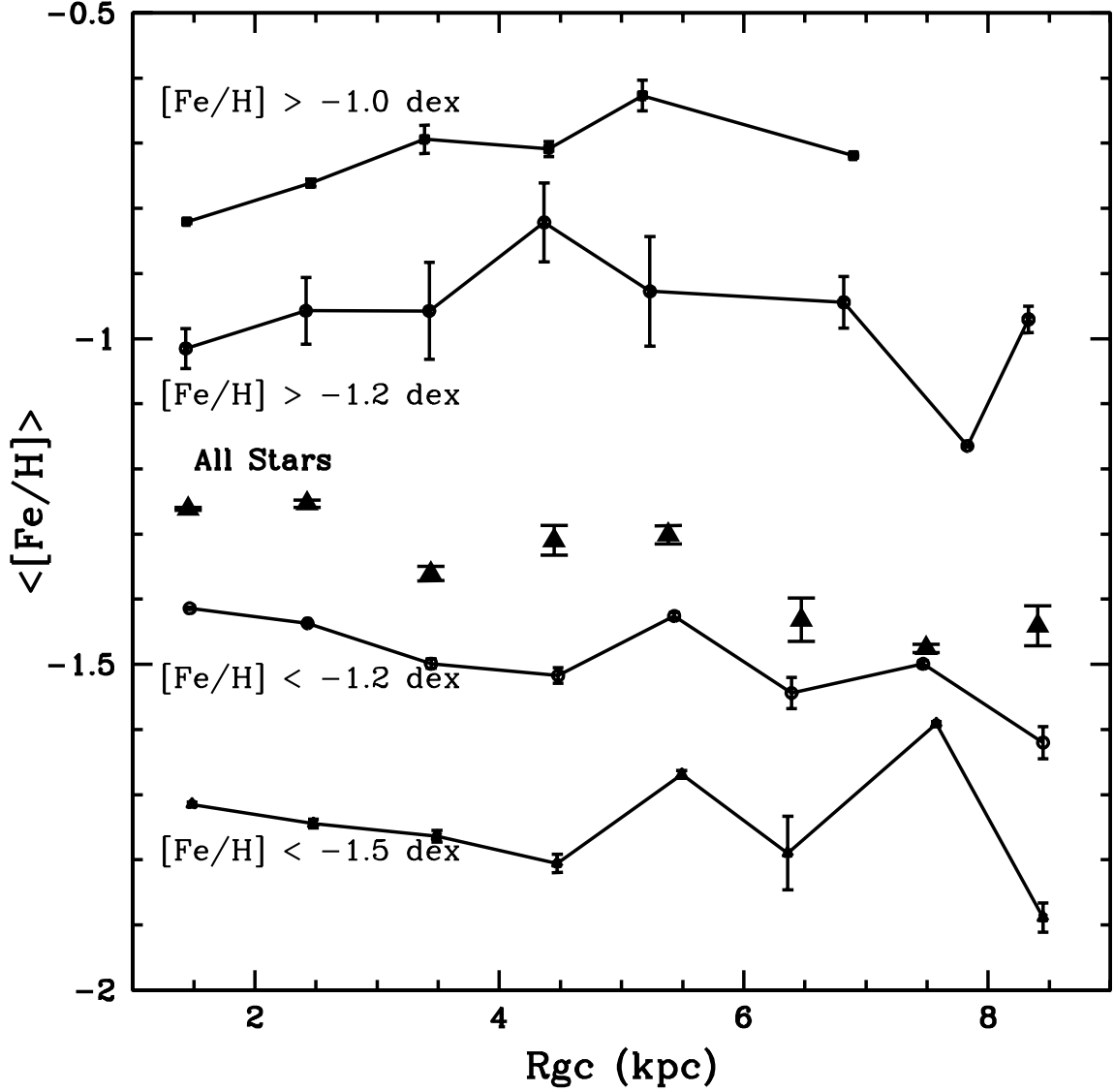}
\caption{The average metallicity of bulge RR0 Lyrae stars in each 1.0 kpc 
bin as a function of galactocentric distance.  The error bars indicate
the error in the mean.
Also shown is the average metallicity of the metal-rich and metal-poor
bulge RR0 Lyrae stars in each 1.0 kpc bin.   
\label{ploteleven}}
\end{figure}

We deredden each star using the minimum-light $(V-R)$ color of each RR0
Lyrae star, and convert color excess to $A_V$ using an $R_{V,VR}$ of 4.3 
\citep{kunder08}. We find the galactocentric distance to each RR0 Lyrae 
star:
\begin{equation}
R_{gc} = 10^{(V-M_V-A_V)/5} \times \sqrt{ (\cos(b)\cos(l)-R_{\odot})^2 + 
(\cos(b)\sin(l))^2 + \sin(b)^2},
\label{distance}
\end{equation}
where $R_{gc}$ is the galactocentric distance in parsecs, $R_{\odot}$ is the 
distance from the Sun to the Galactocenter, and $b$ and $l$ are 
Galactic latitude and longitude, respectively.  The mode of 
our RR0 Lyrae sample is used as $R_{\odot}$.  The average \feh
of all stars lying within a 1.0 kpc bin is calculated, and plotted 
against the galactocentric distance in Figure~\ref{ploteleven}.  
Next, the bulge RR Lyrae stars are separated according to metallicity, 
and the same procedure is implemented for two metal-rich cuts 
($\feh > -$1.0 dex and $\feh > -$1.2 dex) and two metal-poor 
RR Lyrae star cuts ($\feh < -$1.2 dex and $\feh < -$1.5 dex).


There is a correlation of \feh with radial distance depending on
the metallicity of the RR0 Lyrae stars.  At further distances from the bulge, 
the metallicity in general decreases for the metal-poor stars.  
However, this trend is reversed for the metal-rich bulge stars.
Once the metal-rich stars approach $\rm \feh > -$1.5, the 
trend with Galactocentric distance diminishes.  From the spectra of the 
Galactic Bulge red giant population, \citet{minniti96}
found that the velocity dispersion 
of the metal-rich giants decreases with increasing Galactocentric 
distance, whereas the velocity dispersion of the metal poor giants 
shows no trend.
The result obtained here, that the metal-rich and metal-poor RR Lyrae
stars show a different trend in galactocentric distance, is similar in
spirit.

There may be some substructures in this \feh versus radial distance plot.  
Between 4.0 and 6.5 kpc, there is the first ``break'' in a linear metallicity
curve.  By looking at the Cepheid galactic gradient,
\citet{kov05} speculate that near 6 kpc, a transition from the young-disk
population to the outer regions of the Galactic Bulge population occurs.
Perhaps we see such a signature using bulge RR0 Lyrae stars as well.  

It is also interesting to consider if there are any trends with \feh
in the Sgr galaxy.  Coordinate transformations with the origin
at the center of the Sgr galaxy are performed to determine the
distance to each Sgr RR0 Lyrae star from the center of the Sgr galaxy.
We take M54 to be the nucleus of Sgr \citep{monaco05} and 
use \citet{layden00} M54 RR0 Lyrae sample, combined with 
Equation~\ref{distance}, to determine the distance to M54 as 27 kpc.
This distance to the center of Sgr is based on RR0 Lyrae stars 
using the same $M_V-$\feh relationship adopted in this work to
minimize systematic effects.  However, the coordinate transformations 
to the Sgr frame will be somewhat sensitive to the exact value of the 
distance to Sgr's center.  

\begin{figure}[htb]
\includegraphics[width=16cm]{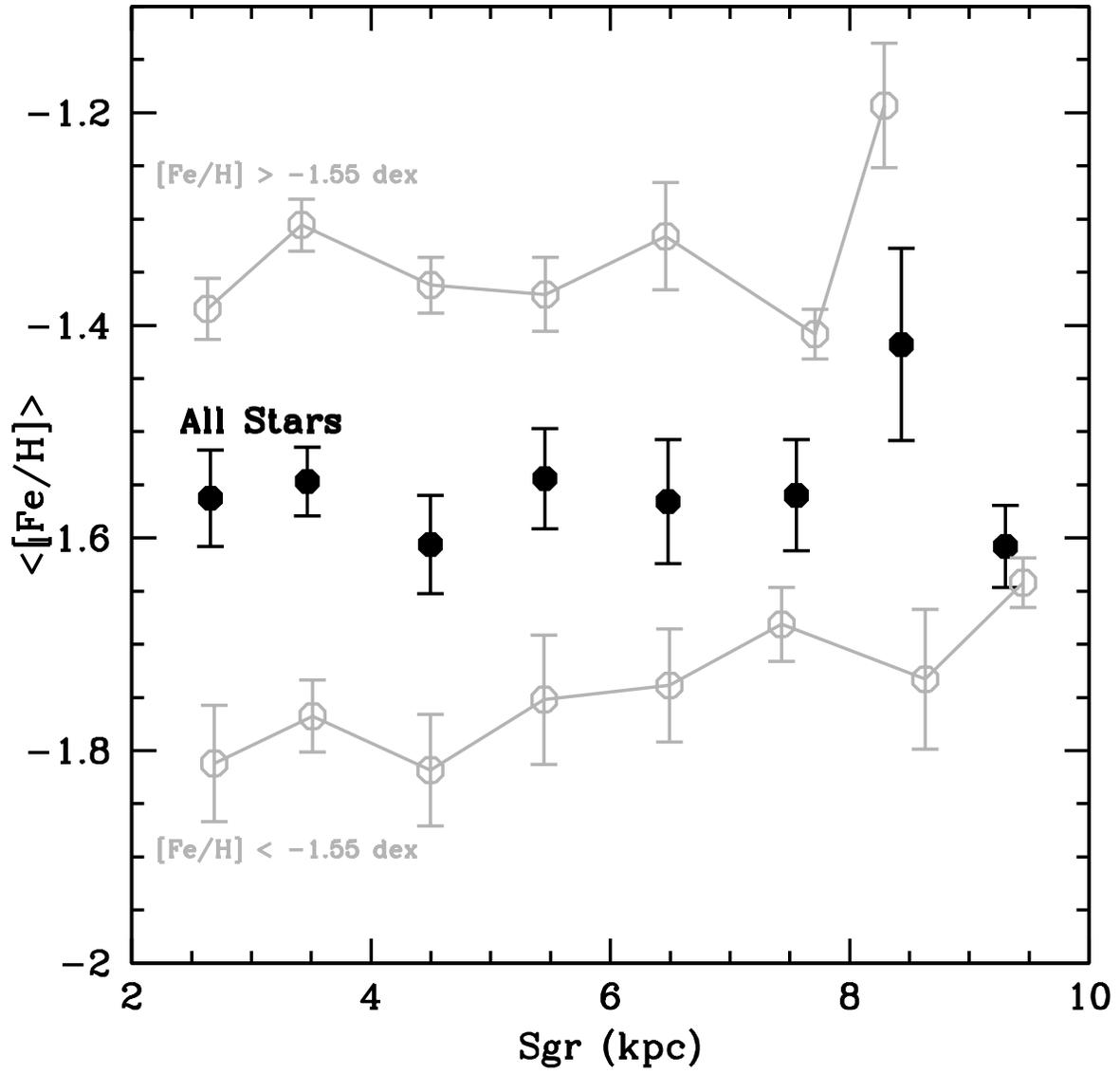}
\caption{Same as Figure~~\ref{plottwelve} but for the Sgr RR0 Lyrae stars.
\label{plottwelve}}
\end{figure}

The average \feh of all Sgr stars lying within a 1.0 kpc bin is calculated 
and plotted against the distance of the center of Sgr in 
Figure~\ref{plottwelve}.
The metal-poor Sgr RR0 Lyrae stars show a slight correlation of 
\feh with radial distance.  On average, the further from the center of
Sgr, the more metal-rich the RR0 Lyrae star.  The metal-rich and
intermediate RR0 Lyrae stars do not display a metallicity gradient.
\citet{cseresnjes01} did not find evidence for a spatial metallicity 
gradient in the RR Lyrae population of Sgr when carrying out a period 
analysis on $\sim$ 3700 Sgr RR Lyrae stars located
just a little south, in galactic coordinates, of the MACHO fields.
Some of the RR Lyrae stars in their study actually overlap with the 
MACHO fields.  

The correlation of \feh with radial distance within the metal-poor
stars is in the opposite direction to the metallicity gradient found 
from giant stars by \citet{ald01, marconi98, bellazzini99}.  However,
metallicity gradients in Sgr may depend on the type of the stellar 
population under investigation.

\section{Conclusion}
Metallicity measurements of 2690 RR0 Lyrae stars are presented 
from the MACHO bulge database using the Fourier
coefficients of their light curves, with an error of 0.20 dex.  
Parameters derived for individual stars are available electronically in
Table~\ref{tab1} for the bulge stars and Table~\ref{tab2} for Sgr stars. 
We find that RR0 stars in the Galactic Bulge span a broad metallicity range 
from $\rm \feh = -2.26$ to $-0.15$ dex, and have an average metallicity of 
$\rm <\feh>$ $= -$1.25.  This compares favorably to other bulge 
metallicity studies.
  
Using mean magnitudes of MACHO RR Lyrae stars we searched for the 
evidence of the Galactic bar, and found a slight signature of
a Galactic bar with Galactic latitude $|b| < 3.5^\circ$.
The most straightforward interpretation of the absence of 
a strong bar is that the metal-rich RR Lyrae in the bulge represent a 
different population than the metal-poor majority.  
As we also found that the average metallicity in this
region is considerably less varied and diverse, we conclude that there is 
evidence indicating the presence of a population belonging to the inner 
extension of the halo, which is relatively metal-poor and which 
could be among the oldest known stars in our Galaxy.  

The RR0 Lyrae stars believed
to be in the northernmost extension of the Sgr galaxy have a 
$\rm <\feh>$ $= -$1.53 which is on average
more metal poor than the RR0 Lyrae stars from the bulge.  This value lies
within the range of metallicity of \citet{marconi98}, and agrees well
with the metallicity found for the $V$-magnitude data of RR0 stars given
by \citet{mateo}.

Based on metallicities of RR0 Lyrae stars, we
estimate their absolute magnitude and distances to all stars.  The distance
to the Galactic Center is 8.0 $\pm$ 0.3 kpc, where the error takes into
account the estimated error in the MACHO photometry calibration as well
as the zero-point in the RR0 Lyrae absolute magnitude relation.
A correlation between metallicity and galactocentric distance is found.
For metal-poor RR Lyrae stars the closer a star is to the Galactic center,
on average, the more metal rich it is.  However for the metal-rich 
RR0 Lyrae stars, we find that this is not the case; instead, on average, 
the closer a star is to the Galactic center, the more metal poor it is.  
The result that the metal-rich and metal-poor RR0 Lyrae stars show a 
different trend in galactocentric distance needs to be addressed when 
modeling the chemical evolution of the Milky Way.

The Sgr RR0 Lyrae stars do not show a strong metallicity gradient.
However, the metal-poor Sgr RR0 Lyrae stars do indicate a metallicity
gradient of $\sim$ +0.15 dex from $\sim$ 2 kpc to 10 kpc of the
galaxy.  (The further the RR0 Lyrae star from the center of Sgr,
the more metal rich.)  This metallicity gradient is in the opposite
direction as that found using giant stars by 
\citet{ald01, marconi98, bellazzini99}.

\acknowledgments

It is our sincere pleasure to bestow thanks to Piotr Popowski, 
Kem Cook, and Sergei Nikolaev for their valuable insight, help and 
encouragement.  Also, thank you to Siobahn Morgan,
Jan Lub, Andy Layden, Alistair Walker and Kenneth Janes for helpful
responses to our e-mail questions.

\appendix
\section{Spectroscopic versus photometric abundances}

Some recent work has suggested that the Fourier parameter-based 
[Fe/H] calibration gives discrepant results, especially in the
low metallicity regime
\citep[see discussions in][]{diFab05, grat04, nemec04}.
In \citet{new}, the iron
abundances of 79 RR0 Lyrae stars derived from the Fourier parameters
were compared with those obtained from low-dispersion spectroscopy.  
It was shown
that almost all discrepant estimates are the results of some defects or
peculiarities either in the photometry or in the spectroscopy (i.e.
spectroscopic abundances based on only two measurements or high noise
in light curve).  


Here a comparison is made between spectroscopic and Fourier
metallicities to asses the accuracy of our \feh determination.  
Such an analysis has been performed by \citet{jk96} and in a variety of
other papers thereafter, \citep[e.g.][]{schwarzenberg98, new} 
the calibration stars used here differs in many respects.  
First, a variety of globular clusters with RR Lyrae lightcurves which have 
previously over over- or under-estimated the published
cluster metallicities when using the \citet{jk96} metallicity formula, 
are addressed.
Second, this set includes MACHO LMC lightcurves with
low resolution spectroscopy in the bar of the Large Magellanic Cloud. 
As a by product of this analysis, the best deviation parameter for our 
particular Fourier decomposition is found.  

\subsection{Star by star comparisons}
\begin{figure}[htb]
\includegraphics[width=16cm]{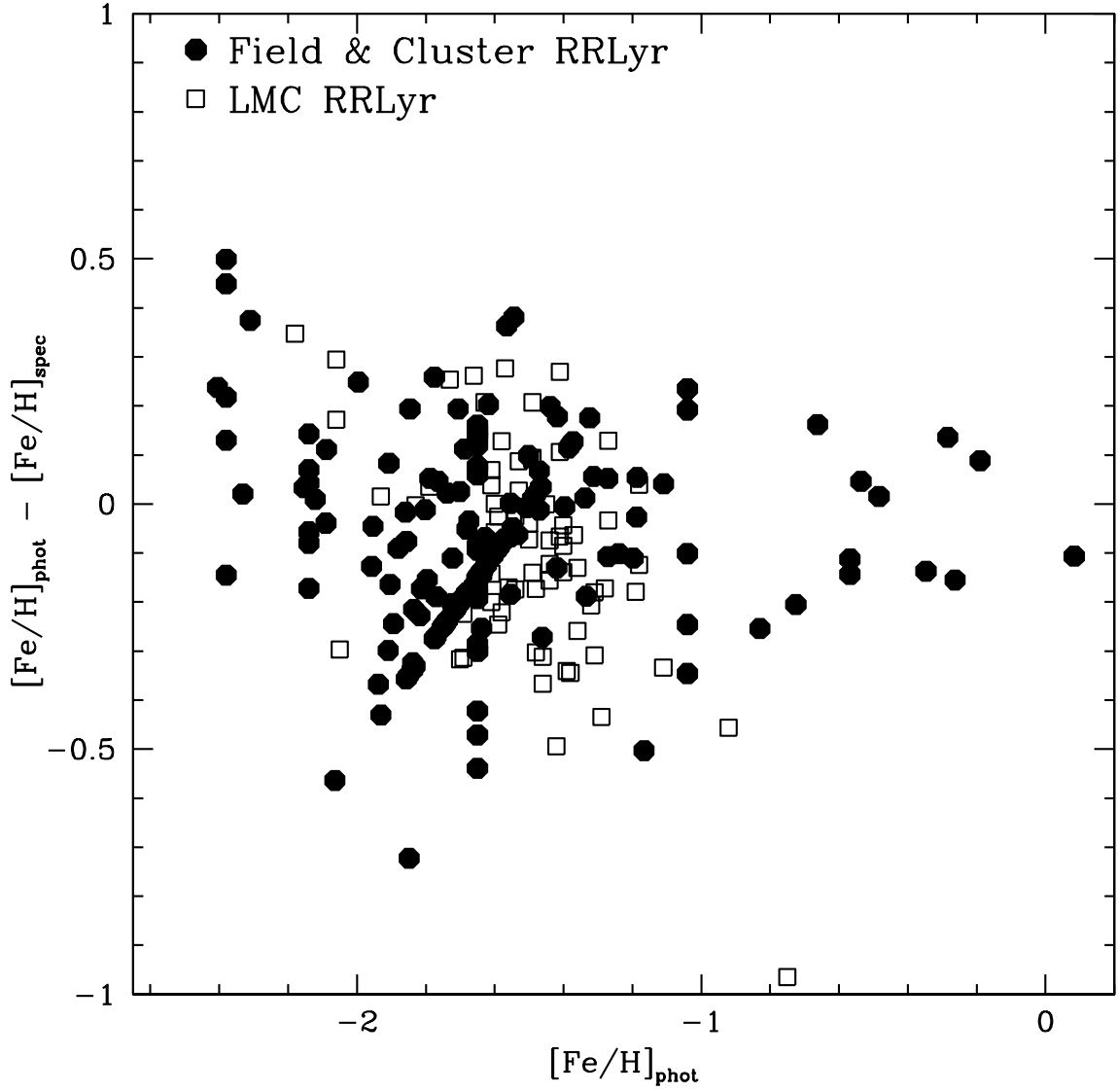}
\caption{The difference between the photometric
and spectroscopic metallicities as a function of the spectroscopic
metallicity.  
\label{plotfourteen}}
\end{figure}

Figure~\ref{plotthirteen} shows a star-by-star comparison of the metallicity 
obtained here photometrically ($\rm \feh_{phot}$) versus
published spectroscopic metallicities ($\rm \feh_{spec}$) of a variety 
of field, cluster, and LMC RR0 Lyrae stars.  The spectroscopic 
metallicities of the 61 RR0 Lyrae field stars 
come from the \citet{jk96} sample and references therein.
The spectroscopic metallicities of the globular clusters 
are discussed in turn below.  The lightcurves of these 
stars come from 
\citet{jk96,bori06,grat04,olech03}, and  Andrew Layden's website
\footnote{\tt http://physics.bgsu.edu/\~layden/ASTRO/DATA/rrl\_data.htm}.
The root mean square residual of the fit is consistent with 0.2 dex 
for RR Lyrae stars with $D_m < 4.5$ and a Fourier decomposition with
a fit order of four.
Figure~\ref{plotfourteen} shows the difference between the photometric
and spectroscopic metallicities as a function of the spectroscopic
metallicity.  Discrepant $\rm \Delta \feh$ values do not seem to
favor a particular metallicity regime.

\subsection{The metal-poor end}
\citet{jk96} acknowledged that the equation overestimated \feh 
for the metal poor clusters M68, M92 and the LMC cluster NGC 1841.  
Some years later, \citet{nemec04} reached a similar conclusion in
an investigation of NGC 5053.  However, in the \citet{jk96} calibration
sample, a significant number of metal poor field RR0 Lyrae stars are
used to test their metallicity formula.  The sample extends to
stars slightly more metal poor than
\feh $\sim$ 2.0 on the \citet{jk96} metallicity scale, 
which corresponds to \feh $\sim$ 2.3 on
the Zinn \& West (1984) metallicity scale.  The metal
poor field RR0 Lyrae stars show no increase in deviation from the
\citet{jk96} period-phase-\feh relation.  Thus the globular cluster
discrepancy is puzzling.

On the \citet{ziwe} scale, the \citet{jk96} photometric metallicity of M92
is \feh$= -$2.28 dex.  Recent high dispersion spectroscopy found that M92 
has \feh = $-$2.38 dex (Kraft \& Ivans 2003, on the Zinn \& West 1984 scale), 
as opposed to the -2.63 dex value assumed in \citet{jk96}.  This is in 
agreement with \citet{jk96}.

Lee, Carney \& Habgood (2005) used high-resolution spectroscopy to find
$\rm \feh = -2.16\pm$0.02 for M68 as opposed to the $\rm \feh -2.33\pm$
0.03 value cited by \citet{jk96} for M68.
This value is closer to the metallicity originally derived by 
\citet{jk96} of $-$2.05 dex (on the Zinn \& West 1984 scale).  
\citet{kraft03} find $\rm \feh = -$2.43 dex.
We perform a fourth, fifth and eighth order Fourier decomposition 
to the M68 RR0 Lyrae star lightcurves, and calculate a deviation parameter.   
The fourth order Fourier decomposition results in five stars with a 
$D_m <$ 4.5, and the average \feh is $\rm -2.16 \pm$ 0.12, in
excellent agreement with the most recent M68 metallicity value.
Increasing the deviation parameter significantly increases the scatter
in average photometric metallicity for this cluster.  
The eighth order Fourier decomposition is similar in result to
the fifth order Fourier decomposition, and yields a \feh value
almost identical to the value of M68 found for M68 in \citet{jk96}
of $-$2.05 dex.  We therefore adopt a fourth order Fourier decomposition 
and a $D_m <$ 4.5 criteria for use in this paper.

NGC 5053 is known to have an extremely low metal abundance of 
$\rm \feh = -$2.4 dex (Kraft \& Ivans 2003, which is based upon
Ca II triplet observations). 
\citet{nemec04} obtained photometry
of 10 RR Lyrae stars in this cluster and used \citet{jk96}
metallicity formula to find their photometric metallicities.  They
find the mean \feh = -2.1 dex on the \citet{ziwe}
scale, about $\sim$ 0.3 dex more metal rich than it's published metallicity.
It is notable, however,
the \citet{nemec04} photometry shows scatter in the phased lightcurve.
Perhaps the difference of the derived photometric metallicities in NGC 5053
is a function of the quality of these lightcurves as opposed to a failure
in the method.

\subsection{The metal-rich end}
The \citet{jk96} calibration sample also contained a significant number of 
metal rich field stars (up to \feh = +0.08 dex).
The metal rich field RR0 Lyrae stars show no increase 
in deviation from the \citet{jk96} period-phase-\feh relation.  However
recent papers have suggested that the RR0 Lyrae stars in the metal
rich clusters of NGC 6441 and NGC 6388, 
($\rm \feh = -0.53 \pm$ 0.11 and $\rm -0.60 \pm$ 0.15, respectively; 
Armandroff \& Zinn 1988), give photometric metallicities
that are much more metal poor than the spectroscopic metallicity of the
cluster \citep{pritzl01, pritzl02}.  Since then, using
the Hubble Space Telescope, \citet{pritzl03}
found that their previous ground based photometry of the NGC 6441
RR Lyrae stars in the center of the cluster is largely affected by 
blends, resulting in photometry errors of up to 1.5 magnitudes in 
the $V$-band.  
Unfortunately, there are not enough $V$-band lightcurve 
points from the \citet{pritzl03} data to derive \feh with a small deviation 
parameter.

There are two \citet{pritzl01} RR0 Lyrae stars in 
NGC 6441 with a $D_m$ $<$ 4.5, and both of these stars give
\feh $\sim$ -1.3 dex.  One of these stars
is only 32'' from the center and therefore is most likely affected by
blending.  
There is one RR0 Lyrae star in the \citet{pritzl02} NGC 6388 
data with a $D_m$ $<$ 4.5.  This star, V22, gives \feh $\sim$ -1.3 dex
for the fairly metal-rich globular cluster of NGC 6388.  
Using low-resolution spectra, \citet{clementini05} finds that
the RR0 Lyrae variables in NGC 6441 range from -0.46 to -1.37 dex.  
This range
does allow for somewhat higher metal abundances of RR0 Lyrae stars
in NGC 6441. 

The RR0 Lyrae variables in NGC 6441 and NGC 6388 tend to have longer 
periods than
a typical RR0 Lyrae star (as long as P = 0.9 d), and no stars with 
periods longer than 0.8 days are used in the \citet{jk96} calibration sample.
Indeed, it is easy
to see the the \citet{jk96} metallicity formula cannot give a metallicity
larger than $-$0.6 dex on the Zinn \& West (1984) scale
for a variable with a period of 0.8 days.  
There are no RR0 Lyrae stars in 
the MACHO sample used in this paper with a period greater than 0.8 days.  
Thus, there are relatively few RR0 stars in our sample with have similar 
properties to those in NGC 6388 and NGC 6441, for which the Fourier 
estimate of [Fe/H] is likely biased.

\subsection{The intermediate metallicity stars}
There is not as much discussion about the validity of the \citet{jk96} 
metallicity formula for the intermediate-metallicity stars, as it
appears to to be robust for these stars.  Using 110 stars from the 
All Sky Automated Survey (ASAS) database, \citet{new} found that 
with the current fitting accuracy 
of 0.17 dex of the Fourier formula, the overall prediction accuracy, 
(i.e., the standard deviation of the predicted \feh) is 0.12 dex. 
This sample does include metal-rich and metal-poor stars, but the majority
of them have an intermediate metallicity, most centered around 
\feh = $-$1.2.  

However, \citet{schwarzenberg98} noted that no obvious
correlation exists between the Butler, Dickens \& Epps (1978) low-dispersion
metallicities of RR0 Lyrae stars in intermediate metallicity cluster
$\omega$ Centauri and their 
photometric metallicities.  Rey et al. (2000), however, did find a correlation
when using metallicities derived from their {\it Caby} photometry.
Recently, Sollima et al. (2006) presented new high-resolution 
spectroscopic metal abundances for 74 RR Lyrae stars in $\omega$ Cen
obtained with FLAMES, 36 of these being RR0 Lyrae stars.  Twenty-
eight of these RR0 Lyrae stars had a $D_m <$ 4.5 and Fourier parameters 
obtained from Olech et al. (2003).
The Olech et al. (2003) Fourier parameters come from light curves
obtained by the Cluster AgeS 
Experiment (CASE) project and have between 200 - 600 observations per
light curve.
The photometric and spectroscopic metallicities of the 28 RR0 Lyrae stars,
which range from $\sim -$2.1 to $-$1.2 dex, 
are in excellent agreement, exhibiting a 0.16 dex dispersion around the line of
unity.

\subsection{MACHO lightcurves}
The MACHO survey used non-standard filters which are converted from 
MACHO instrumental magnitudes $b_M$ and $r_M$, to Johnson $V$.  An error 
in the conversion process in these bands would lead to the incorrect 
shape of the lightcurve.  This in turn would affect the values of 
Fourier coefficients and lead to an incorrect estimate of metallicity.
MACHO LMC RR0 Lyrae stars can be used to address this concern.

\begin{figure}[htb]
\includegraphics[width=16cm]{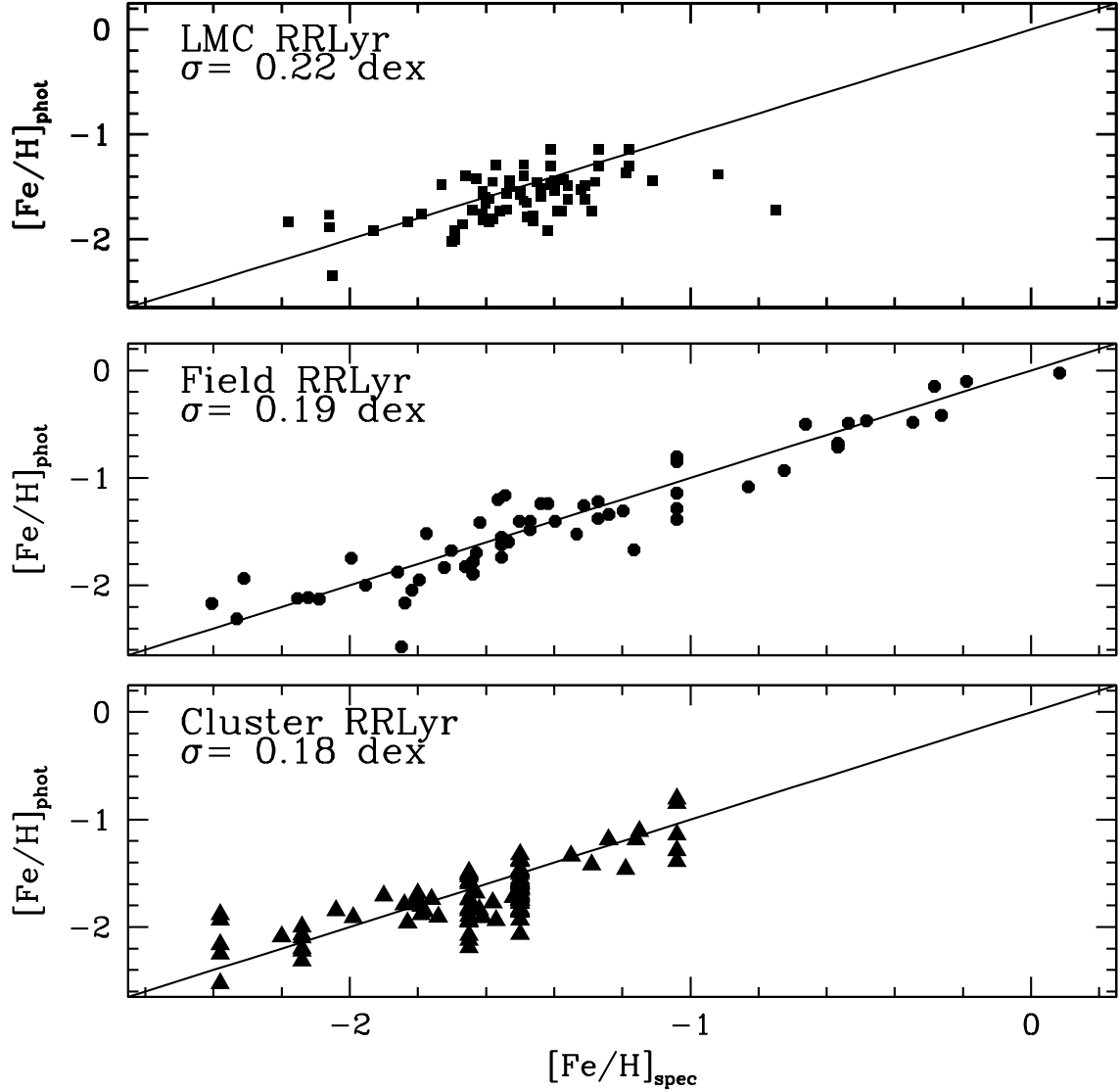}
\caption{A star-by-star comparison of the metallicity obtained photometrically
versus earlier published spectroscopic
metallicities, using RR0 Lyrae stars from the LMC, the field, and from
the globular clusters discussed in the text of Appendix A.
The solid line indicates the line of unity.
\label{plotthirteen}}
\end{figure}

\citet{grat04} find \feh metallicities for 98 RR Lyrae from low
resolution spectroscopy in the bar of the Large Magellanic Cloud with
typical errors of $\pm$ 0.17 dex.  Using their positions in the sky 
and their periods, forty-two of these stars are found to match with 
a MACHO LMC RR0 Lyrae stars from \citet{alc03} .  Borissova et al. (2006) 
measured spectroscopic \feh metallicities 
\citep[in a similar manner to][]{grat04}, 
for $\sim$ 50 MACHO RR Lyrae stars in the LMC.  
Fifteen of these stars are included in the published \citet{alc03} sample.
The \citet{grat04} and \citet{bori06} metallicities are on the 
\citet{harris96} metallicity scale and are shifted by 0.06 dex
to also place them on the \citet{ziwe} scale \citep{grat04}.  
The MACHO photometry is used to determine \feh.
All 15 RR0 Lyrae stars from \citet{bori06} and 38 RR0 Lyrae
stars from \citet{grat04} have a $D_m <$ 4.5.  
Thus, there are 53 LMC stars with spectroscopic \feh abundances that also have
MACHO lightcurves.  The top plot in Figure~\ref{plotthirteen} shows a 
star-by-star comparison of the MACHO photometric LMC metallicities and the 
earlier published spectroscopic metallicities.  The fit is a linear with
a reduced mean scatter (rms) of 0.22 dex around the line of unity, with
an average difference of 0.21 dex.
Increasing the order of the Fourier series does not show any significant 
changes of the (lower) Fourier coefficients when using the MACHO data.  

\citet{diFab05} use 29 RR0 Lyrae
stars to derive the average difference between their photometric and 
spectroscopic metallicities to be 0.3 dex.  As this difference is larger
than what we obtained in the previous paragraph, we match 20 light curves
in the \citet{diFab05} sample with MACHO LMC lightcurves stars that also
have \citet{grat04} metallicities.  A photometric metallicity is
determined for these 20 stars using both the \citet{diFab05} light curves
and the MACHO LMC light curves.  The rms about the line of unity
 is 0.34 dex between the spectroscopic and 
the \citet{diFab05} photometric metallicities, similar to the result found
by \citet{diFab05}.   However, when the MACHO lightcurves are used, 
the rms scatter is reduced to 0.22 dex between the spectroscopic and 
MACHO photometric metallicities, which is in line with our estimated
error.  This is most likely due to the fact the MACHO 
light curves have 7-9 times as many data points as the 
\citet{diFab05} light curves.

\clearpage

\begin{table}
\begin{scriptsize}
\centering
\caption{Bulge RR0 Lyrae Star Metallicities}
\label{tab1}
\begin{tabular}{lcccccccccc} \\ \hline
Star ID & Gal $l$ & Gal $b$ & Period & $V$ mag & $R$ mag & $\feh$ & $D_m$ \\ \hline
101.20648.645 & 3.286 & -2.904 & 0.59006304 & 17.58 & 16.77 & -1.14 & 1.47\\
101.20649.500 & 3.357 & -2.820 & 0.57013899 & 17.11 & 16.46 & -1.34 & 3.78\\
101.20650.531 & 3.385 & -2.821 & 0.55754501 & 17.29 & 16.44 & -1.57 & 3.09\\
101.20650.965 & 3.363 & -2.829 & 0.50520098 & 18.38 & 17.57 & -1.40 & 0.65\\
101.20653.526 & 3.597 & -2.702 & 0.46992201 & 17.26 & 16.53 & -1.04 & 4.42\\
\hline
\end{tabular}
\end{scriptsize}
\end{table}

\begin{table}
\begin{scriptsize}
\centering
\caption{Sgr RR0 Lyrae Star Metallicities}
\label{tab2}
\begin{tabular}{lcccccccccc} \\ \hline
Star ID & Gal $l$ & Gal $b$ & Period & $V$ mag & $R$ mag & $\feh$ & $D_m$ \\ \hline
101.21175.1970 & 3.799 & -2.887 & 0.56146103 & 19.25 & 18.55 & -1.64 & 2.91\\
102.22720.1340 & 3.341 & -4.066 & 0.70013201 & 19.05 & 18.38 & -2.27 & 3.85\\
104.20510.6338 & 2.773 & -3.113 & 0.57012099 & 19.29 & 18.52 & -1.44 & 2.65\\
110.22187.2215 & 2.450 & -4.291 & 0.64345598 & 19.48 & 18.75 & -1.70 & 2.67\\
110.23227.1734 & 2.734 & -4.749 & 0.54225904 & 19.24 & 18.57 & -1.66 & 1.84\\
\hline
\end{tabular}
\end{scriptsize}
\end{table}

\clearpage

\clearpage

\clearpage

\clearpage

\end{document}